\documentclass[twocolumn,twocolappendix]{aastex63}
\usepackage[utf8]{inputenc}
\usepackage{enumitem}
\usepackage{graphicx}
\usepackage{hyperref}
\usepackage{mathtools}
\usepackage{comment}
\usepackage{amsmath}
\usepackage{ amssymb }
\usepackage{lipsum} 
\usepackage{float}
\usepackage{longtable}
\usepackage{cleveref}
\usepackage[caption=false]{subfig}
\usepackage{placeins}

\begin{document}
\title{The Statistics of Extended Debris Disks Measured with {\it Gaia} and {\it Planck}}
\author{Jacob Nibauer}\email{jnibauer@sas.upenn.edu}
\affiliation{Center for Particle Cosmology, Department of Physics and Astronomy,\\ University of Pennsylvania, Philadelphia, PA 19104, USA}

\author{Eric Baxter} 
\affiliation{Center for Particle Cosmology, Department of Physics and Astronomy,\\ University of Pennsylvania, Philadelphia, PA 19104, USA}
\affiliation{Kavli Institute for Cosmology, Institute of Astronomy, \\ University of Cambridge, Cambridge CB3 0HA, UK}

\author{Bhuvnesh Jain}
\affiliation{Center for Particle Cosmology, Department of Physics and Astronomy,\\ University of Pennsylvania, Philadelphia, PA 19104, USA}

\begin{abstract}
Thermal emission from debris disks around stars has been measured using targeted and resolved observations.
We present an alternative, likelihood-based approach in which temperature maps from the {\it Planck} CMB survey at 857 and 545 GHz are analyzed in conjunction with stellar positions from {\it Gaia} to estimate the fraction of stars hosting disks and the thermal emission from the disks. The debris disks are not resolved (or even necessarily detected individually) but their statistical properties and the correlations with stellar properties are measured for several thousand stars. We compare our findings with higher sensitivity surveys of smaller samples of stars. For dimmer stars, in particular K and M-dwarfs, we find about 10 percent of stars within 80 pc have emission consistent with debris disks. We also report on 80 candidate disks, the majority of which are not previously identified.   We have previously constrained the properties of Exo-Oort clouds using {\it Planck} data -- with future CMB  surveys both components can be measured for different stellar types, providing a new avenue to study the outer parts of planetary systems. 
\vspace{2cm}
\end{abstract} 

\section{Introduction}

Many stars are known to be surrounded by orbiting disks of dust and debris.  The sizes of these debris disks range from tens to hundreds of AU.  Their presence is thought to proceed from the protoplanetary disk phase of a newly formed star, with dust grains originating primarily from the collisional cascade of planetesmials \citep{2008ARA&A..46..339W}.

Absorption of optical light from the parent star by the debris disk heats the disk to temperatures on the order of tens to a hundred Kelvin, depending on the stellar type and the distance from the star.  Given their temperatures and large surface areas, debris disks can be strong emitters in the infrared and submillimeter.  Observations in the infrared probe the inner, hotter parts of the disks, and small grains.  Observations in the submillimeter, however, are better suited to studying extended debris disks and larger grains. Submillimeter observations of debris disks have been somewhat limited, but serve as a powerful probe to understanding debris disk properties around low luminosity stars \citep{Lestrade_2006}. In this work, we focus on observations in the submillimeter because of the availability of deep, large-area submillimeter maps from {\it Planck} \citep{2018arXiv180706205P}. However, the methodology that we develop could in principle be applied to observations at any frequency.

Targeted, high-resolution observations of debris disks in the submillimeter have long been used to study these objects.   Observations with ALMA at 223~GHz, for instance, can resolve the structure of nearby disks such as that of {\it Fomalhaut} \citep{MacGregor:2017}. 

In this analysis, we take a complementary approach to studying debris disks, using low-resolution, but large-area data from {\it Planck}.  Using the measured positions and distances of {\it Gaia}-detected stars \citep{2018A&A...616A..11G,2018A&A...616A...1G}, we constrain the parameters of simple disk models.  While the {\it Planck} observations cannot resolve features in individual debris disks, they can be used instead to constrain the statistics of the emission from these objects and correlations with, for instance, stellar properties. Open questions in this area include the incidence of debris disks around stars of different spectral types, and the number of dim, low mass stars hosting cool debris disks \citep{doi:10.1146/annurev-astro-081817-052035,Lestrade_2009,Binks_2017}. 

Several previous works have studied the statistics of debris disk populations.  For instance,  \citet{Patel:2014,Patel:2017} used infrared observations with the all-sky WISE survey to identify stars with excess thermal emission beyond the stellar photosphere. Studies of debris disks with wide field surveys provide measurements of a substantially larger number of potential debris disk hosting stars than pointed observations. Other all-sky surveys used in previous debris disk analyses include IRAS, 2MASS, and AKARI, all observing at shorter wavelengths compared to {\it Planck} \citep{Rhee_2007,2017A&A...601A..72I}.

Our analysis is unique in several respects relative to  previous analyses.  For one, we use submillimeter observations from {\it Planck}, relying primarily on the 545 and 857~GHz bands. We are therefore more sensitive to colder debris disks, the outskirts of debris disks, and larger grains. An ancillary advantage of these low-frequency observations is that the signal from the stellar photosphere is expected to be negligible.  For observations with e.g. WISE, on the other hand, emission from the star itself may be non-negligible \citep{Patel:2014}.  Another novel feature of our analysis is that we do not require any individual debris disk to be detected in the {\it Planck} maps.  Since we combine constraints from many possible debris disks, disks with signals too weak to be detected individually in the {\it Planck} maps can still contribute to our constraints.  

We note that our  analysis is sensitive to other sources of emission correlated with stars in addition to that from debris disks.  For instance, protoplanetary disks \citep{Williams:2011} or exo-Oort clouds \citep{Baxter:2018} could also source similar emission signals.  Our results should in some sense be viewed as constraining all of these potential sources.  However, for the present analysis, we expect debris disk emission to dominate over other sources.

The paper is organized as follows.   In \S\ref{sec:modeling} we describe the ingredients of our debris disk model; in \S\ref{sec:data} we describe the data from {\it Planck} and {\it Gaia} that we use to constrain this model; in \S\ref{sec:methodology} we describe how we process the data; in \S\ref{sec:simulations} we describe the application of our analysis to simulated data sets; in \S\ref{sec:results} we present our results on actual data, and we conclude in \S\ref{sec:discussion}.

\section{Modeling}\label{sec:modeling}

We use {\it Planck} data to constrain the parameters of a model describing the population of extended debris disks around stars detected by {\it Gaia}.  We do not require that any individual disk be detected in the {\it Planck} maps\footnote{Note that some nearby disks, such as that of {\it Fomalhaut}, can be detected with signal-to-noise greater than one in the {\it Planck} maps, see e.g. \citet{Baxter:2018}.}. 

Debris disk temperatures are typically in the range of 30 to 100 K towards the outskirts of the disk, depending on the distance from the parent star.  These temperatures correspond to Wien peaks at roughly 1800 and 6000~GHz, beyond the frequency range probed by {\it Planck}.  We therefore rely on the highest frequency channels from {\it Planck}, namely those at 545 and 857~GHz.  Relative to debris disk observations with e.g. WISE or CHARA \citep{2017A&A...608A.113N}, we expect our analysis with {\it Planck} to be more sensitive to extended disks or disks around less luminous stars.  

The {\it Planck} 545 and 857~GHz channels have approximate beam sizes of 4.7 and 4.3 arcminutes FWHM, respectively.  Since the typical debris disk radius ranges from $\sim$ 30--300AU  \citep{Pawellek:2014}, essentially all debris disks beyond a few parsecs from Earth will be unresolved by {\it Planck}.  Therefore, to a very good approximation, we can measure only one number for each disk, namely the amplitude of its emission signal.  We refer to the amplitude 
measurement for the $i$th star as $d_{i}$, which we measure as a surface brightness with units of ${\rm Jy}/{\rm sr}$ (i.e. the units of the {\it Planck} 545 and 857~GHz maps).  We will discuss how $d_i$ is obtained from the {\it Planck} maps in \S\ref{sec:data_and_background_measurement}.  

Our model for $d_i$, which we denote with $\hat{d}_i$, is the sum of a (possibly zero) debris disk contribution, and a term accounting for background emission from other sources:
\begin{equation}\label{eq:d_i}
\hat{d}_i = \alpha_i  \frac{1}{A_{\rm eff} } \frac{L_\nu}{4\pi r_i^2} + B_i.
\end{equation}
The quantity $\alpha_i$ controls the presence or absence of a debris disk, with $\alpha_i = 1$ if the star has a debris disk and $\alpha_i = 0$ otherwise.   $L_\nu$ is the luminosity of the debris disk signal in the {\it Planck} band of interest and $r_i$ is the distance to the $i^{th}$ star.  $A_{\rm eff}$ is an effective sky area (in steradians) that captures both the impact of the {\it Planck} beam and impact of the chosen map pixelization; we will discuss this factor in more detail in \S\ref{sec:Eff Area}. Finally, $B_i$ describes the contribution of backgrounds and noise sources to the measurement of $d_i$.  We simplify our notation by defining
\begin{equation}\label{eq: s definition}
s \equiv  \frac{L_\nu}{4\pi A_{\rm  eff}},
\end{equation}
so that Eq.~\ref{eq:d_i} becomes
\begin{equation}
   \hat{d}_i = \alpha_i \frac{s}{ r_i^2} + B_i.
\end{equation}

At the frequencies of interest and for the majority of stars, emission from a possible debris disk will completely dominate over any emission from the stellar photosphere because of the much larger solid angle subtended by the disk relative to the star.  We therefore ignore possible photospheric emission from the star for the majority of this analysis, but discuss this possible contribution in \S\ref{sec:discussion}. 

Assuming contributions to $B_i$ from the star itself are negligible,  $B_i$ is then dominated by extragalactic backgrounds and galactic dust. It is therefore reasonable to assume that the background distribution is independent of the signal distribution.  However, this assumption could be broken if debris disks are more likely to form in places in the galaxy with significant nearby dust, for example.  

Given the assumption of independent signal and backgrounds, we can write the likelihood of the data in the case that $\alpha_i = 1$ as
\begin{equation}
P(d_i|\theta, \alpha_i = 1) = \int ds \, P_s(s) P_B(d_i - s/r_i^2),
\end{equation}
where $\theta$ represents the model parameters describing the debris disk emission, $P_s(s)$ characterizes the probability distribution of $s$, and $P_B(B)$ characterizes the probability distribution of the backgrounds.  On the other hand, in the absence of a debris disk (i.e. $\alpha_i = 0$), the probability of measuring $d_i$ is just given by the background distribution:
\begin{equation}
P(d_i|\theta, \alpha_i = 0) = P_B(d_i).
\end{equation}

Since we do not expect to get many high signal-to-noise detections of debris disks, we are most interested in the case where we cannot determine whether $\alpha_i= 0$ or $\alpha_i = 1$ at high significance.  We therefore marginalize over $\alpha_i$ to obtain the likelihood: 
\begin{multline}\label{eq:likelihood}
    P\left(d_i | \theta \right) = \int  P_B \left(d_i - s/r_i^2\right)P_{s}\left(s\right)P\left(\alpha_i=1\right) ds \\
    + P_B\left(d_i\right)P\left(\alpha_i = 0 \right).
\end{multline}
We will return to the possibility of measuring $\alpha_i$ for individual stars in \S\ref{sec:candidates}.  

We suppose for simplicity that all stars (after imposing some selection criteria to be discussed below) have the same probability of hosting a debris disk. Under this assumption, we define the disk fraction 
\begin{equation}
    f_{\rm disk} \equiv P\left(\alpha_i = 1\right).
\end{equation} 
It follows that $P\left(\alpha_i = 0\right) = 1-f_{\rm disk}$. 

In order to compute the likelihood defined above, we must assume some form for the background distribution $P_B(B)$ and the signal distribution $P_s(s)$.  As we discuss in more detail in \S\ref{sec:data_and_background_measurement}, we find that we can accurately model the background fluctuations in the ${\it Planck}$ maps with a Gaussian model with mean $\mu_{B,i}$ and variance $\sigma_{B,i}^2$, where these two parameters vary  from star to star (with index $i$ labelling the different  stars).   Eq.~\ref{eq:likelihood} then becomes
\begin{multline}\label{eq:likelihood_model}
    P\left(d_i|\theta\right) = \\
    f_{\rm disk} \int\limits_{0}^{\infty} \mathcal{N} (d_i  - \frac{s}{r_i^2} | \mu = \mu_{B,i},\sigma^2 = \sigma_{B,i}^2) P_s\left(s\right)ds  \\
    + (1-f_{\rm disk})\mathcal{N}\left(d_i | \mu = \mu_{B,i},\sigma^2 = \sigma_{B,i}^2\right),
\end{multline}
where $\mathcal{N}(x| \mu, \sigma^2)$ is the normal distribution.

We now consider the signal distribution, $P_s(s)$, for which we will explore two models.  Perhaps the simplest possible model is that all debris disks have the same luminosity so that 
\begin{eqnarray}
\label{eq:constant_lum}
P_s(s) = \delta(s - s_d),
\end{eqnarray}
where $\delta$ is a Dirac $\delta$-function, and $s_d$  describes the luminosity of all disks.   In this case, the model parameters are $\theta = \{f_{\rm disk},s_d\}$.  We will refer to this as the constant luminosity model.  

While the constant luminosity model has the advantage of simplicity, it is unlikely to reflect the true distribution of debris disk properties in nature.  For instance, recent collisional events of planetesimals are thought to result in an extended tail of strong emitters \citep{Bryden:2006}.  A common choice for modeling the distribution of debris disk luminosities is a log-normal model \citep[e.g.][]{Bryden:2006}.  Since the total debris disk luminosity can be reasonably expected to scale with the luminosity of the parent star, it is common in the literature to work with the fractional luminosity, i.e. the ratio of the bolometric disk luminosity to that of the parent star.  For simplicity, we will instead adopt a log-normal model for $s$ itself.  In this case, we have
\begin{equation}\label{eq:lognormal}
   \small{
      P_{s}\left(s|\mu_{\ln{s}},\sigma_{\ln{s}}^2\right) = \frac{1}{s\sqrt{2\pi}\sigma_{\ln{s}}}\exp\Bigl[{-\frac{
    (\ln{s} - \mu_{\ln{s}})^2}{2\sigma_{\ln{s}}^2}}\Bigr]}.
\end{equation}  
For this model, the parameters are  $\theta = \{f,\mu_{\ln{s}},\sigma_{\ln{s}}^2\} $.

Finally, we assume that the $d_i$ measured for each star is independent of the measurements for other stars, allowing us to write the total likelihood as
\begin{equation}\label{eq:total_likelihood}
   \mathcal{L}\left(\{d_i\}|\theta\right) =  \prod_i P\left(d_i|\theta \right),
\end{equation}
where $\{d_i\}$ represents the set of measurements for all stars, and the product runs over all stars.  The assumption of independent measurements may be invalidated by correlated background fluctuations for nearby stars.  However, our tests on simulated data suggest that this effect does not lead to a bias in our  constraints (see \S\ref{sec:simulations}).  The likelihood in Eq.~\ref{eq:total_likelihood} is computed by substituting either Eq.~\ref{eq:constant_lum} (constant luminosity model) or Eq.~\ref{eq:lognormal} (lognormal model) into Eq.~\ref{eq:likelihood_model} and Eq.~\ref{eq:total_likelihood}.

\section{Data}
\label{sec:data}

\subsection{{\it Planck} data}\label{sec:planck_data}

The ${\it Planck}$ satellite observed the full sky in nine frequency bands from $30~\rm{GHz}$ to $857~\rm{GHz}$ \citep{2018arXiv180706205P}. Our analysis relies on the $545$ and $857~\rm{GHz}$ bands, as these frequencies are most well matched with the emission of possible debris disks, and are also the highest resolution bands. We use the publicly available maps at \url{https://pla.esac.esa.int/}.  These maps are downgraded to a \texttt{Healpix} \citep{2005ApJ...622..759G} resolution of  $N_{\rm side} = 1024$ for the final analysis in order to better match the pixel size to the beam size.

\subsection{{\it Gaia} data}\label{sec:gaia_data}

The stellar positions and distances  used in this analysis are taken from the \textit{Gaia} DR2 release \citep{Gaia:2018}. This data release contains parallaxes for stars down to a limiting magnitude of G=21. The {\it Gaia} satellite also obtains low-resolution spectroscopy, from which broad-band photometry is extracted and stellar parameters are estimated \citep{Andrae:2018}.  We use use the {\it Gaia} color measurements to impose various cuts  on the stellar catalog (see  \S\ref{sec: Star Selection}), and focus on nearby stars with $d < 100\,{\rm pc}$. The {\it Gaia} data set is virtually complete for main sequence stars in this volume, and the expected parallax errors are essentially negligible for our purposes. 

\subsection{Masking}\label{sec:mask}

\begin{figure}[t]
    \includegraphics[width=8.5cm]{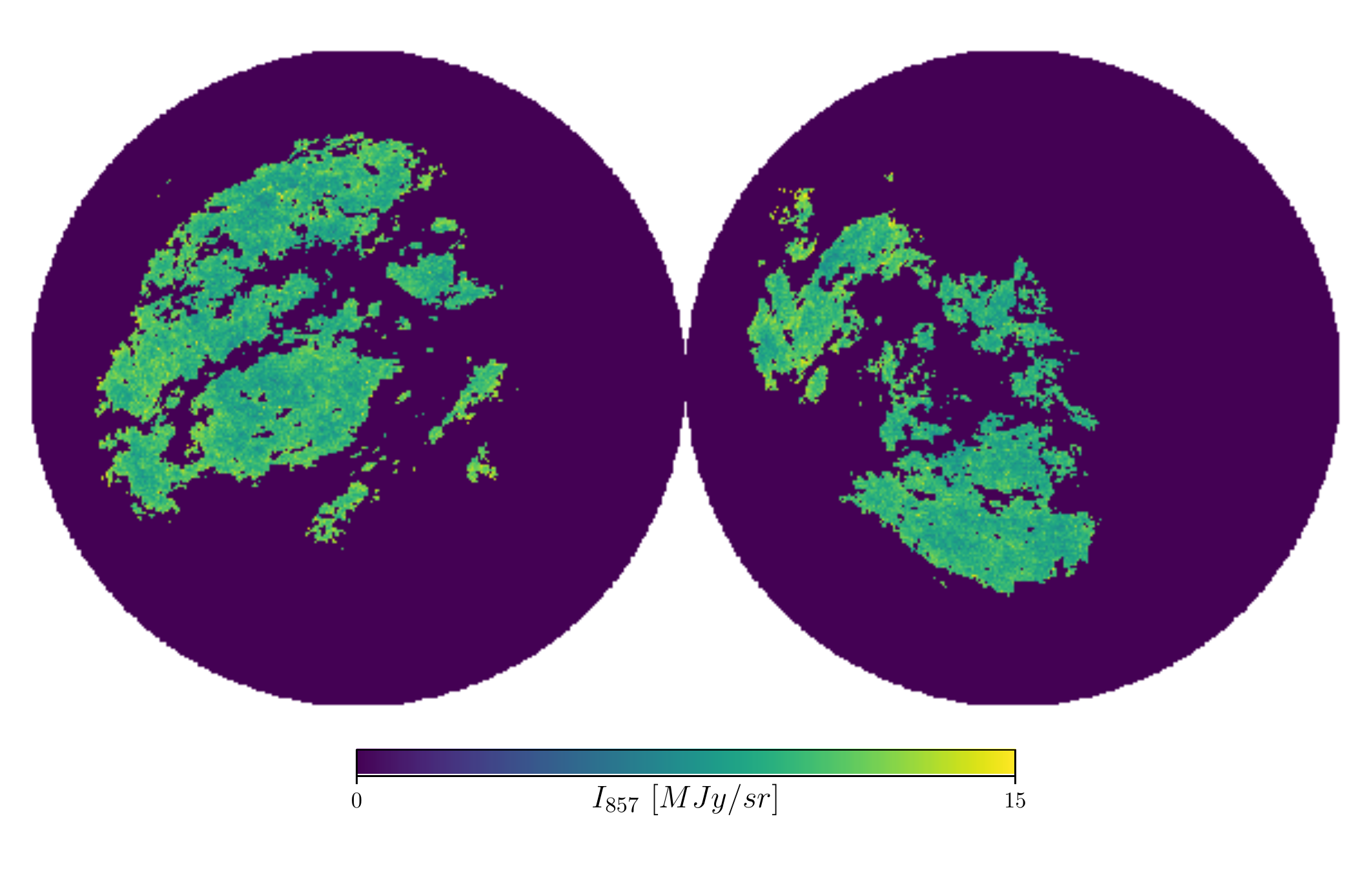}
    \centering
    \caption{Orthographic projection showing the north and south poles of the {\it Planck} map at $857~{\rm GHz}$ after imposing point source masking and the HI mask.  The HI mask removes 90\% of the sky with the highest HI column density. Note that orthographic projections are not area preserving. }\label{fig:Planck Masked}
\end{figure}

Galactic backgrounds and non-debris disk point sources are a potential challenge for our analysis.  The high density of {\it Gaia} stars across the sky means that it is not unlikely to have a chance alignment between a star and a bright background fluctuation or extragalactic source.  We reduce the impact of such background fluctuations by imposing a conservative sky mask. The mask is derived from a map of  HI column density by \citet{Lenz:2017}, which has been shown to correlate strongly with thermal emission from galactic cirrus at low HI column density.  In our fiducial analysis, we leave unmasked any pixels whose column densities are in the bottom 10th percentile of this map (see Fig. \ref{fig:Planck Masked}). Additionally, we use the PSCz catalog from \cite{Saunders:2000} to mask IRAS-detected galaxies with an aperture of 5' and NGC objects with a radius of 15'. 

\begin{figure}[h]
    \includegraphics[width=9cm]{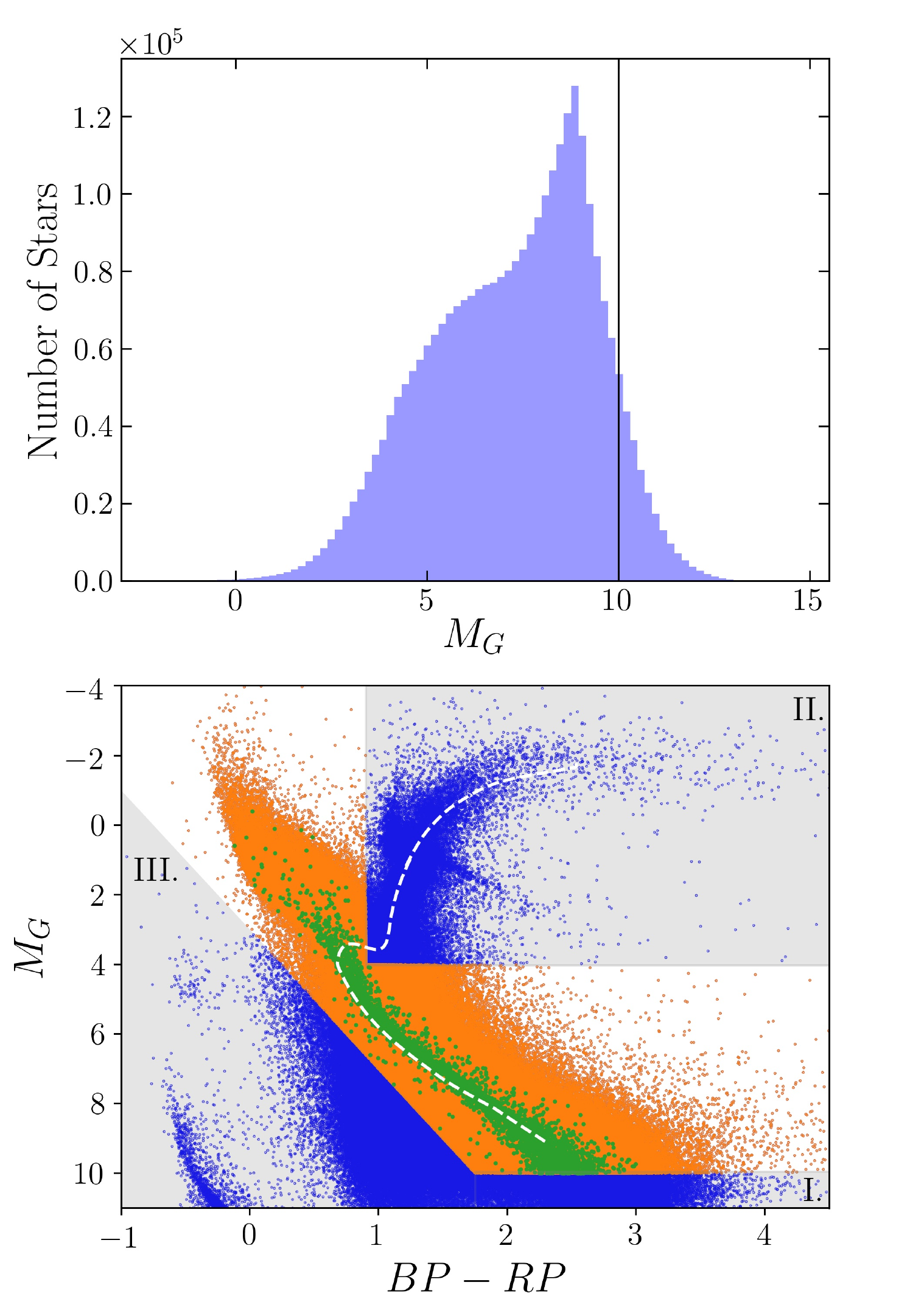}
    \centering
    \caption{The selection of stars from {\it Gaia} used in our analysis. \textbf{Top:} The distribution of {\it Gaia} G band absolute magnitudes for main sequence stars in our sample. The steep fall off at $M_G \approx 9$ is the result of {\it Gaia's} stellar selection. The vertical line at $M_G = 10$ marks the maximum $M_G$ for stars included in our analysis. \textbf{Bottom:} Orange points indicate stars that survive our baseline masking, while blue points are {\it Gaia} stars that are cut from our analysis. Green points indicate stars in our fiducial sample, while the white dotted line is a main sequence isochrone with solar-like parameters. We note that our sample includes a high number of faint, cool stars with $M_G > M_{G,\rm{Sun}}$.}\label{fig:M_G Distribution and HR Diagram}
\end{figure}

\section{Methodology}
\label{sec:methodology}

\subsection{Measurement of $d_i$ and background model}
\label{sec:data_and_background_measurement}

Since {\it Planck} does not resolve the debris disks, we can only measure a single amplitude value for each star, $d_i$.  In the interest of simplicity, our estimator for $d_i$ is just the pixel value of the {\it Planck} map at a resolution $N_{\rm side} = 1024$ for the pixel in which the star resides.  Such a pixel is roughly $3.4$~arcmin on a side, which can be compared to the $4.3$~arcmin FWHM of the {\it Planck} beam at 857~GHz.   Consequently, for a debris disk located in the center of a pixel, the pixel will contain a majority of the debris disk flux.  We correct for missing flux due to the beam and pixelization as described in \S\ref{sec:Eff Area}.

In principle, one could construct an estimator for $d_i$ that uses the known position of the star inside the pixel and the beam shape to obtain a more optimal estimate of the signal from the star.  However, we find that the simple technique adopted here performs well in simulations, and is significantly simpler and faster than fitting a beam template or applying a matched filter to the maps.  We postpone a more optimal analysis to future work.  

The distribution of the backgrounds in the {\it Planck} 545 and 857~GHz maps varies significantly across the sky, as these maps receive large contributions from galactic cirrus and other non-isotropic sources.  To form an estimate of the local background distribution for each star, we draw an annulus around the star --- from $\theta_{\rm min}$ to $\theta_{\rm max}$ --- and measure the mean, $\hat{\mu}_{B,i}$, and variance, $\hat{\sigma}_{B,i}^2$, of the map within this annulus.  Our estimate of the background contribution to $d_i$ is simply a normal distribution with this mean and variance.

\subsection{Effective area, $A_{\rm eff}$}\label{sec:Eff Area}

We now describe the computation of the effective area, $A_{\rm eff}$ in Eq.~\ref{eq:d_i}. This factor effectively measures the relation between flux from a point source, and surface brightness measured in the {\it Planck} maps.  By definition,  a point source with flux $X$~Jy will on average lead to a pixel with surface brightness $(X/A_{\rm eff})\,{\rm Jy}/{\rm sr}$. $A_{\rm eff}$ therefore includes both the impact of the instrument beam, and the impact of pixelization of the beam-smoothed map.  We estimate $A_{\rm eff}$ with a Monte Carlo procedure, where we first choose random positions in an $N_{\rm side} = 2048$ map, add a signal with known flux to the pixel containing the random position, convolve the resultant map with the {\it Planck} beam (using a Gaussian approximation to the beam) and downgrade it to the $N_{\rm side} = 1024$ resolution of the actual data analysis.  Finally, the surface brightness values in the pixels of the degraded map centered on the input positions are computed, and the ratio is taken relative to the input signal.  We repeat this process many times to build statistics.  We find $A_{\rm eff} = 31.6\pm 0.003\ {\rm arcmin}^2$.

\subsection{Star selection}\label{sec: Star Selection}

Debris disks have been measured around a wide variety of stars, though they have been most commonly detected in spectral classes A-K \citep{Krivov:2010}. Because of this, we focus our analysis on main sequence stars in spectral classes A-M. We choose not to omit M stars from our baseline analysis, given that submillimeter surveys such as {\it Planck} are well suited to measuring debris disks around a cooler sample of stars \citep{Lestrade_2006}.
We do, however, remove giant stars from our sample as they can be strong emitters in the wavebands of interest due to gas and dust ejecta  \citep{2018MNRAS.475.2282V}. Additionally, we remove possible white dwarfs, as they have been found to have low debris disk detection rates \citep{2012ApJ...760...26B}. These selections are accomplished with several magnitude and color cuts:
\begin{enumerate}[label=\Roman*.]
    \item We remove stars with absolute magnitude in the {\it Gaia} G band, $M_G$, greater than 10.
    
    \item Stars with $M_G<4$ and color $BP-RP > 0.9$ are removed as possible giants.
    
    \item Stars with $M_G>4\times(BP-RP)+3$ are removed as possible white dwarfs.
\end{enumerate}
Our baseline cuts are illustrated by the orange points in the bottom panel of Fig. \ref{fig:M_G Distribution and HR Diagram}, and leave roughly $2.5 \times 10^6$ stars across the full sky out to 330~pc. The roman numerals in the shaded regions correspond to the three cuts described above, while the green points indicate stars in our fiducial sample (see \S\ref{sec:planck_sim_results}). We include a main sequence isochrone \citep{2013A&A...553A..62V,2012A&A...547A...5V} to aid in visually distinguishing a star from the main sequence.

A unique feature of our stellar population compared to that of previous debris disk analyses is the higher density of dim, low mass stars in our sample. The top panel of Fig.~\ref{fig:M_G Distribution and HR Diagram} illustrates the distribution of {\it Gaia} G band absolute magnitudes after performing the main sequence selection described above. The step falloff in the number of stars at $M_G \sim 9$ is a consequence of Gaia's stellar selection. While we explore removing dim stars from our analysis in \S\ref{sec: changing star selection}, our baseline stellar population is clearly dominated by stars with $M_G > 7$. As illustrated in Appendix \ref{app:spec_type} we find that $\sim 75\%$ of the main sequence stars in our sample are in spectral classes K-M for varying distance cuts. 

\section{Analysis of simulated data}\label{sec:simulations}

\subsection{Generating simulations}

We test our analysis pipelines and model assumptions by generating and analyzing simulated {\it Planck} maps with mock debris disk signals.

To generate these simulated data sets, we start by randomly generating mock debris disk luminosities using either the constant luminosity or log-normal models. The amplitude of the signals in either case is chosen to roughly match (on average) measurements of Fomalhaut's debris disk at 857~GHz by  \citet{Herschel_Fomalhaut}.  For the constant luminosity model, we set $s_d \approx 2.96 \times10^{40} {\rm \ W Hz^{-1} Sr^{-1}}$.  We note that Fomalhaut is a mid to large size debris disk.  We assume $f_{\rm disk} = 0.15$ following \citet{Krivov:2010}, which found that approximately 15\% of stars in spectral classes A-K host debris disks.  The mock debris disk emission is then randomly turned off with probability $1-f_{\rm disk}$.  Next, the mock disks are assigned distances drawn from the true distribution of distances to {\it Gaia} stars and inserted into \texttt{Healpix} maps at randomly located pixels of an $N_{\rm side} = 2048$ map.  This map is then convolved with the {\it Planck} beam and degraded to a resolution of $N_{\rm side} = 1024$.  In order to increase the statistical precision of the simulation tests, we generate the simulations using a density of stars five times greater than found in the actual data.

The background contributions in the simulated maps are generated either as Gaussian noise or by using the {\it Planck} maps themselves. In the Gaussian case, we choose a noise level comparable to the variance of the {\it Planck} maps over the unmasked region. This model is a poor imitation of the highly non-Gaussian galactic backgrounds, but it useful for our purposes because in this case we can know the background distribution perfectly. On the other hand, using the {\it Planck} maps themselves as an estimate of the background is justified since the debris disk emission is expected to make a small contribution to these maps. When using the {\it Planck} maps as background estimates, we only place simulated debris disks within pixels that do not host actual {\it Gaia} stars out to three time the maximum distance. 

\subsection{Gaussian simulation results}\label{sec:gauss_sim_results}

We now discuss the results of analyzing the simulated {\it Planck} maps, starting with the Gaussian simulations.  Given that the Gaussian simulations are very unlike the real data, they are most useful for the purposes of validating our pipelines; we therefore relegate most discussion of these results  to Appendix~\ref{app: simulations}.  For the Gaussian background case, our model is a complete description of the data and should therefore recover an unbiased estimate of the input parameters.  The results of this analysis are shown in Fig.~\ref{fig:Gaussian Simulation}. We find that the input parameters are recovered to within the uncertainties.  

As the distance cut is increased, the number of stars increases. At the same time, however, the average flux from a star will decrease as the distance cut is increased.  Consequently, if the background model is correct, one expects the constraints to asymptote to a constant level of uncertainty as the distance cut is increased (see Appendix~\ref{app:snr} for a more detailed justification of this point). This behavior in the limit of a large distance cut is illustrated by the dashed contours in Fig.~\ref{fig:Gaussian Simulation}, where we use a 250~pc maximum distance cut.

Using the Gaussian simulations, we find that if there are multiple disks in the same pixel, the parameter constraints can be biased. Additionally, neighboring pixels containing disks also lead to bias in simulations, since, when convolving with the {\it Planck} beam, the signal from nearby sources combine. Our modeling does not account for these features, so we impose the restriction that no star in a pixel can be within three times the maximum distance cut of another star in the same pixel, and no star included in the analysis can be within 3~arcmin of another star in the catalog.  This restriction eliminates bias in the simulations, but could conceivably impose a selection bias in the actual analysis, as stars with nearby neighbors may have differing debris disk properties. For a maximum distance cut less than 100~pc, these restrictions typically remove about $20-30\%$ of stars that survive the HI masking (see \S\ref{sec:mask}).   

\subsection{{\it Planck} simulation results}\label{sec:planck_sim_results}
\begin{figure*}[ht]
    \includegraphics[width=18.5cm]{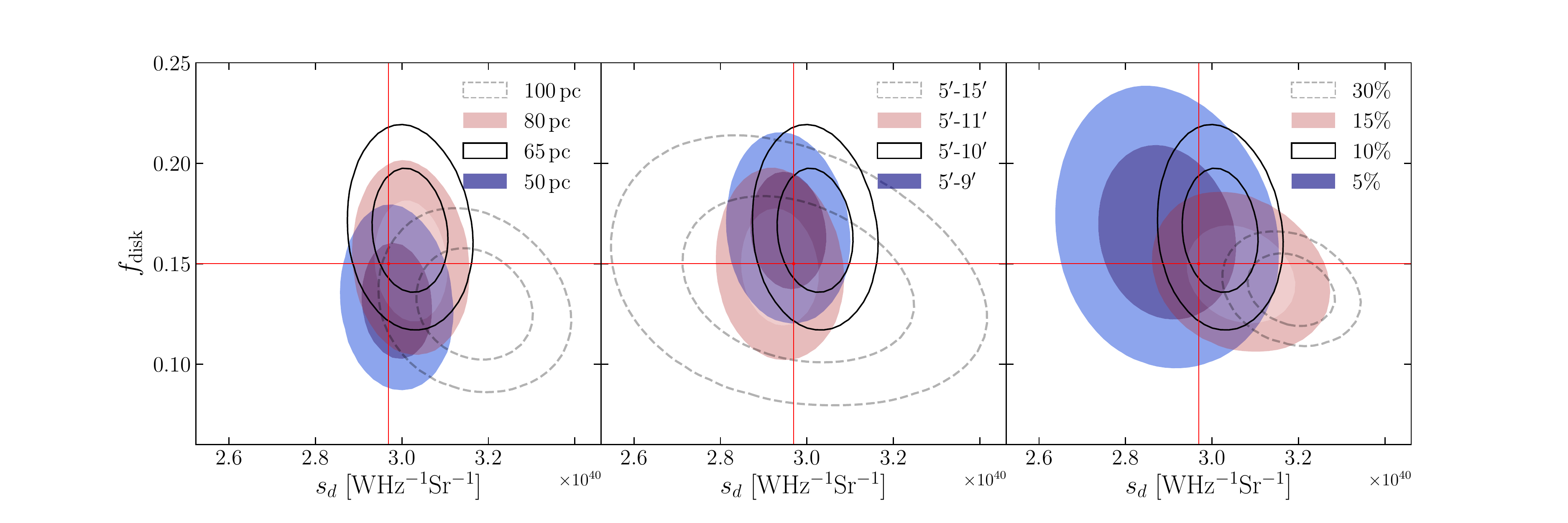}
    \centering
    \caption{Results of simulated analyses for {\it Planck} backgrounds and constant luminosity model. For each realization, we use 5 times the number of stars found in real data.  Input parameter values are shown with the red crosshair; inner  and  outer contours represent 68 and 95\% confidence regions,  respectively.  \textbf{Left:}  results for varying distance cuts, where the number of stars is 5000, 9800, and 17300 for 50~pc, 65~pc, and 80~pc respectively. \textbf{Middle:} results  for varying annulus width. For each trial, $\theta_{\rm{min}}=5 \ \rm{arcmin}$ while we vary $\theta_{\rm{max}}$. \textbf{Right:} results for Varying HI mask, leaving unmasked regions of the sky with HI column densities in the bottom 5\%, 10\%, 15\%, and 30\%. We find that for reasonable variations around the fiducial analysis choices (shown as unfilled black contours in each panel), the input parameters are recovered without bias.}\label{fig:Constant Luminosity Three Panel}
\end{figure*}

We next analyze the simulations in which the background model is the {\it Planck} maps themselves.  Unlike the Gaussian background case, using the {\it Planck} maps means that our Gaussian approximation to the local background distribution near a disk is not necessarily accurate.  The {\it Planck}-based simulations therefore enable us to test the extent to which our background model can yield accurate parameter constraints.

The results of analyzing the {\it  Planck} background simulations are shown in Fig.~\ref{fig:Constant Luminosity Three Panel} for the constant luminosity model, and Fig.~\ref{fig:Planck Simulation Lognormal} for the log-normal luminosity model. In the case of the constant luminosity model, we choose the same input parameters as used in the Gaussian simulations (\S\ref{sec:gauss_sim_results}). For the log-normal luminosity model, we choose a fiducial value of $\sigma_{\ln{s}} = 1$, and select the value of $\mu_{\ln{s}}$ that corresponds to our earlier choice of $s_d$ in the Gaussian simulations. As before, we set $f_{\rm{disk}} = 0.15$. 

For both tests, we consider three types of variations around our fiducial analysis choices: (1) using different stellar distance cuts, (2) using different annulus sizes for the purposes of estimating local backgrounds, (3) using different masking thresholds for the HI mask.  Our fiducial choices include a distance cut of $65$~pc, an annulus size of $\theta_{\rm{min}} = 5 \ {\rm{arcmin}}$ and $\theta_{\rm{max}} = 10 \ {\rm{arcmin}}$, and an HI mask of 10\%. 

\begin{figure*}
    \includegraphics[width=14cm]{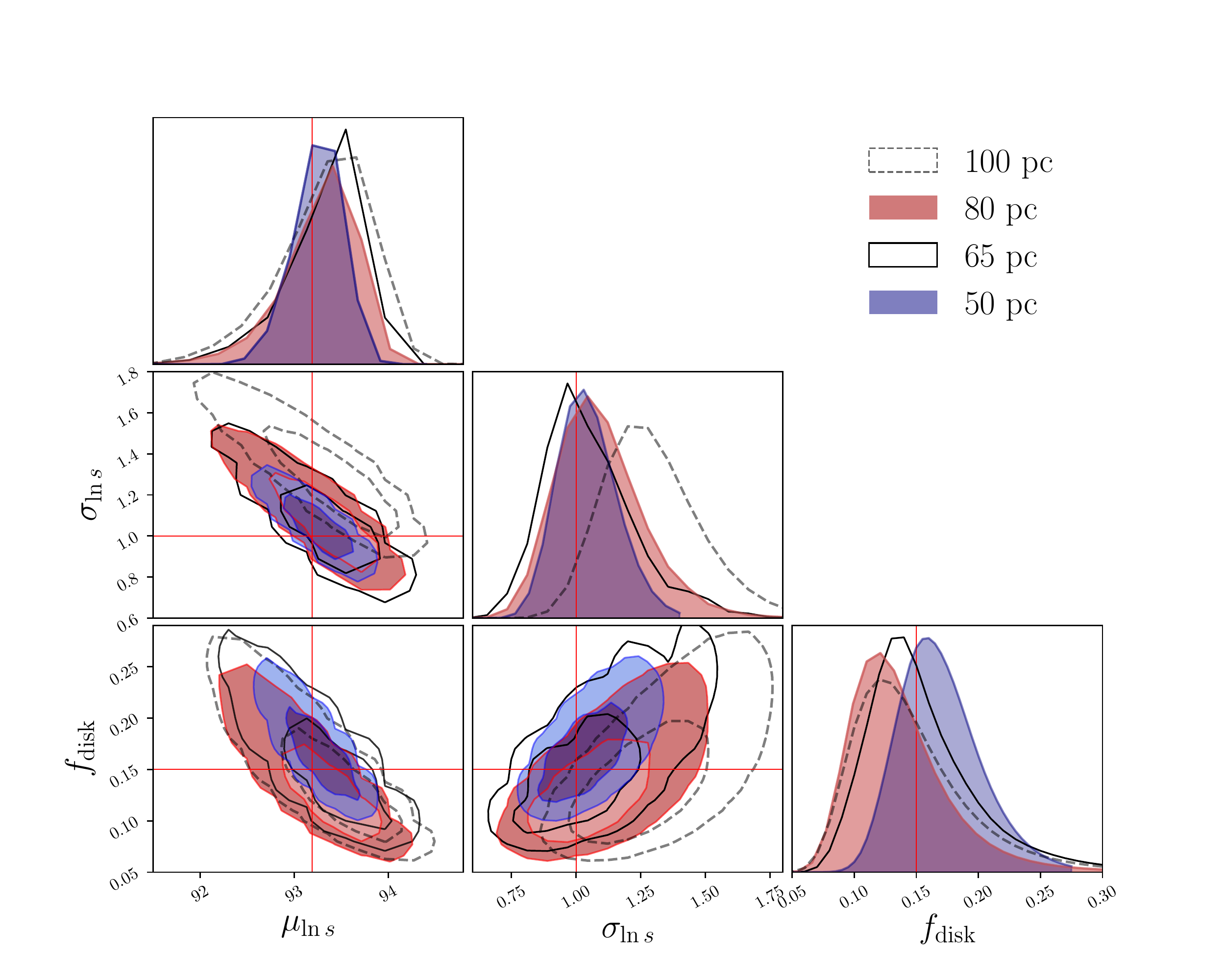}
    \centering
    \caption{{\it Planck} background simulation results for the log-normal luminosity model. In each panel, the input parameters are depicted by red lines. As in Fig.~\ref{fig:Constant Luminosity Three Panel}, we use 5 times the number of stars found in real data for each realization. Results shown are for a varying distance cut with a fixed HI mask of 10\% and annulus width of 5 arcmin. In  Fig.~\ref{fig:Planck Simulation Lognormal}, we present additional log-normal luminosity model results for variations around the fiducial annulus size and HI mask. The fiducial choice in all log-normal simulation plots is depicted as the unfilled solid line contours. We find that all reasonable variations around the fiducial choices lead to the successful recovery of the input parameters within the uncertainties. }\label{fig:Planck Simulation Lognormal Distance Panel}
\end{figure*}

For distant stars, for which the expected signal will be less than the backgrounds, inaccuracies of our background model can lead to parameter biases.  Since the number of stars increases rapidly with distance cut, using a large distance cut could conceivably lead to a bias in the recovered constraints.  We therefore ensure that reasonable variations around our fiducial distance cut do not lead to biases in the simulations. The results for this test are shown in the left panel of Fig.~\ref{fig:Constant Luminosity Three Panel} and in Fig.~\ref{fig:Planck Simulation Lognormal Distance Panel} for the constant luminosity and log-normal models respectively.  We find that small variations around the fiducial 65~pc cut do not lead to biases. In the limit of a large distance cut, however, the inaccuracies of our Gaussian background model become apparent. 

Next, we consider the impact of varying the annulus sizes used to estimate the background distribution near each star. The middle panels of Fig.~\ref{fig:Constant Luminosity Three Panel} and Fig.~\ref{fig:Planck Simulation Lognormal} show that for reasonable variations around our fiducial choice, the size of the annulus does not significantly impact our results. We find, however, that if the annulus size is increased significantly beyond the fiducial choice, the variance estimate of the background will also increase  (since the annulus will then sample regions with different levels of galactic dust).  In this case, the uncertainties on the model parameters increase, and the estimate of the signal flux will be driven to zero. This is illustrated by the dashed contours in the middle panel of both figures.

The impact of varying the HI mask is shown in the right panel of Fig.~\ref{fig:Constant Luminosity Three Panel} and Fig.~\ref{fig:Planck Simulation Lognormal}. More conservative masks reduce the impact of non-isotropic and non-Gaussian backgrounds, at the cost of decreasing the number of stars. Significantly less conservative HI masks are found to lead to potential biases, as emission from non-debris disk sources becomes more common.  

We also consider the bias that can result from a mismatch between the true and assumed debris disk luminosity  distributions.  In particular,  we analyze simulations for which the disks have been generated assuming the log-normal model, but which we analyze using the constant luminosity model. The results of this test are shown in Fig.~\ref{fig:ConstLum_withLogNorm}. The constant luminosity model returns increasingly biased results as the width of the log-normal distribution is increased. We also overplot a curve corresponding to fixed total luminosity, i.e. fixed $f_{\rm disk} L_{\nu}$. As the width of the log-normal is increased, the model fits the high brightness fluctuations by increasing $L_{\nu}$ while decreasing $f_{\rm disk}$, keeping the total luminosity fixed to match the fixed mean brightness of the log-normal distribution.

\section{Results with data}
\label{sec:results}

We now present the results of our analysis applied to the {\it Planck} data, using the maps at both 545 and 857~GHz.  

\subsection{Null test}
\label{sec:null_test}

We first consider the results of our analysis applied to pixels on the sky that are not hosts to {\it Gaia}-detected stars. This test should yield no detection of signal (i.e. be consistent with $f_{\rm disk} = 0$).  The results of the null tests are show in Appendix~\ref{app: Results with Data}. For both the constant luminosity and log-normal models, the null tests yield non-detections given our fiducial analysis choices.

We note that prior to applying our conservative HI mask, null tests would occasionally fail.  Inspection of the data points leading to the failing null tests revealed that these were often driven by extended patches of dust emission, or by apparent extragalactic sources.  After imposing the conservative HI mask and the extragalatic source mask discussed in \S\ref{sec:planck_sim_results}, we find that the null tests pass across many realizations.  The imposition of the conservative masks effectively ensures that the backgrounds are more closely described by the Gaussian background model that we have adopted.

\subsection{Constant luminosity model results}\label{sec:Const_Lum Results}

We first consider the results of analyzing the data assuming the constant luminosity model.  As noted above, if the underlying debris disk population follows a log-normal distribution of luminosities (as we suspect it does), assuming a constant luminosity model can result in significant biases.  We therefore relegate the constant luminosity model results to Appendix \ref{app:Constant Lum}.  

\subsection{Log-normal model applied to known debris disks}\label{sec:Log Normal Model on Known Disks}

\begin{figure*}[t]
    \includegraphics[width=18.5cm]{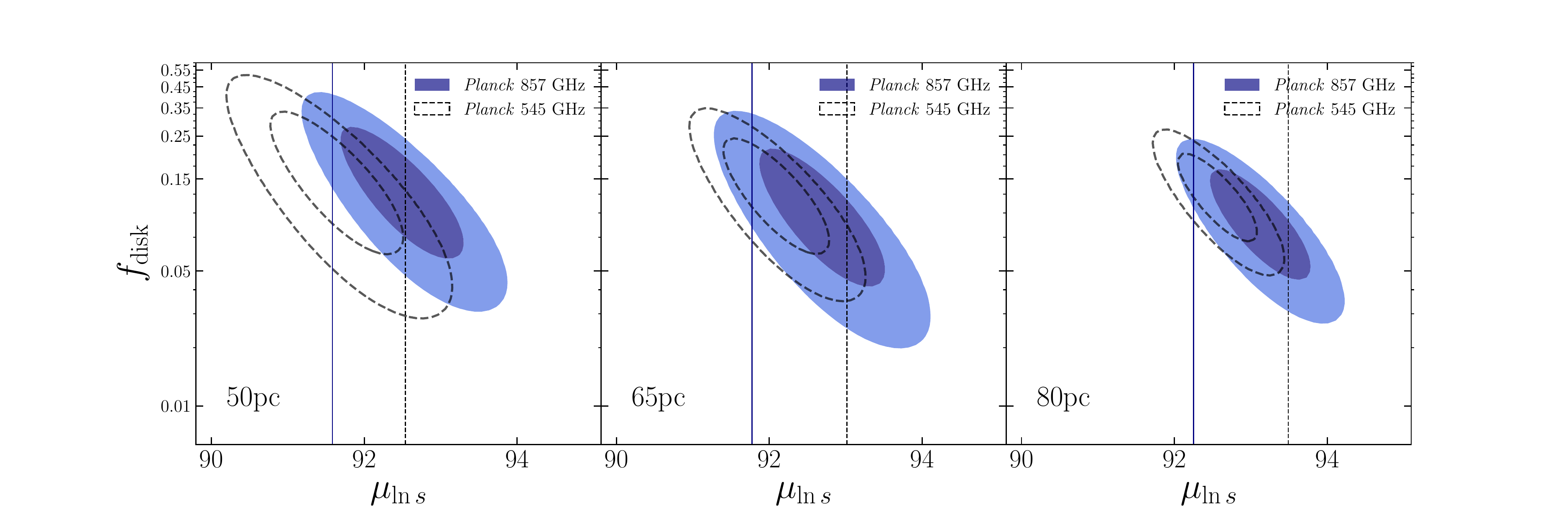}
    \centering
    \caption{Log-normal luminosity model results from the {\it Planck} 545 and 857~GHz channels for a fixed value of $\sigma_{\ln{s}} = 1.5$. We use the fiducial analysis choices, including an HI mask leaving 10\% of the sky uncovered and an annulus size with $\theta_{\rm{min}} = 5, \ \theta_{\rm{max}} = 10$ arcmin. The 68\% and 95\% confidence regions are shown for three distance cuts. The number of {\it Gaia} stars included in the analysis is $\sim 1000, 2000, \ \rm{and} \ 3500$ for the 50~pc, 65~pc, and 80~pc cuts respectively. For each distance cut, we find statistically significant results at greater than $4\sigma$ from $f_{\rm{disk}} = 0$. In each panel, the vertical blue line corresponds to the expected Rayleigh-Jeans scaling in $\mu_{\ln{s}}$, based on the best fit $\mu_{\ln{s}}$ in the 857~GHz channel. The vertical dashed line in each panel illustrates the same expected shift, but based on the best fit $\mu_{\ln{s}}$ in the 545~GHz channel.
    }\label{fig:LogNorm Fixed Sigma Baseline Results}
\end{figure*}

As motivated above, the log-normal luminosity model is likely a good description of the underlying debris disk luminosity distribution.  We test this model by applying it to a catalog of 75 known debris disks\footnote{\url{https://www.circumstellardisks.org}}. In this case, we expect to recover $f_{\rm disk} = 1$.

The result of this analysis is shown in Fig.~\ref{fig:75 Known Disks}. We find that the constraints are indeed consistent with $f_{\rm disk} = 1$. Fixing $f_{\rm disk}$ to $1$ and marginalizing over $\mu_{\ln{s}}$, we find the 68\% confidence interval on $\sigma_{\ln s}$ to be $[1.29,2.95]$.  

\subsection{Log-normal luminosity model results}\label{sec:Log-Normal Results}

We now apply the log-normal analysis to the full catalog of {\it Gaia}-detected stars.

As before, our fiducial analysis choices are to use a 65~pc maximum distance cut, an HI mask leaving 10\% of the sky uncovered, and to remove all stars with a {\it Gaia} measured G band absolute magnitude, $M_G$, less than 10. We explore how the debris disk constraints vary as a function of $M_G$ in \S\ref{sec: changing star selection}. 

Given the fairly low signal-to-noise of our measurements, we fix $\sigma_{\ln{s}} = 1.5$ in the subsequent analysis.  This value is consistent with the results obtained from our analysis on known debris disks in \S\ref{sec:Log Normal Model on Known Disks}.  We fix $\sigma_{\ln s}$ for two primary reasons.  First, for small values of $\sigma_{\ln{s}}$, the log-normal luminosity model converges to the constant luminosity model, which we find is prone to biases if the true luminosity distribution has significant spread. This is a probable scenario, for a population of debris disks is likely to have an extended tail of strong emitters due to recent collisional events \citep{Bryden:2006}.  Second, for large values of $\sigma_{\ln{s}}$, the log-normal model is more susceptible to fitting background fluctuations in the {\it Planck} maps.

Results for the log-normal luminosity model with $\sigma_{\ln{s}} = 1.5$ are shown in Fig.~\ref{fig:LogNorm Fixed Sigma Baseline Results}. For the fiducial analysis choices, we find constraints (95\% confidence level):
\begin{equation*}
    \begin{split}
    \mu_{\ln{s}} & = 92.5 \pm^{1.04}_{1.02}\\
    f_{\rm{disk}} & = 0.10 \pm^{0.17}_{0.07}.
    \end{split}
\end{equation*}
We determine the statistical significance of our measurements by using a likelihood ratio test to compare the best fit model to the model consistent with no excess signal ($f_{\rm{disk}} = 0$). For our fiducial analysis choices, we find a detection at $4.2\sigma$ with the 857~GHz channel. For the 50~pc and 80~pc distance cuts also presented in Fig.~\ref{fig:LogNorm Fixed Sigma Baseline Results}, we find $6.0\sigma$ and $5.2\sigma$ detections respectively. In the 545~GHz channel, we find even higher statistical significance at 6.6$\sigma$ for the fiducial 65~pc maximum distance cut, $5.6\sigma$ and $8.8\sigma$ for the 50 and 80~pc distance cuts respectively. 

The vertical lines in each plot of Fig.~\ref{fig:LogNorm Fixed Sigma Baseline Results} represents the expected shift in spectral luminosity between frequency channels for a black-body emitter well approximated by the Rayleigh-Jeans law. The position of the blue vertical line is determined by taking the $L_{\nu}$ corresponding to the best fit $\mu_{\ln{s}}$ in the 857~GHz channel, and multiplying by the square ratio of the frequencies, $(545/857)^2$. The position of the dashed lines are determined in a similar way, but by using the best fit $\mu_{\ln{s}}$ from the 545~GHz channel, then applying the appropriate Rayleigh-Jeans scaling. We find that this scaling matches the results in both frequency channels, suggesting that our constraints are dominated by thermal emitters.  This differs from the results found in \S\ref{sec:Const_Lum Results} with the constant luminosity model, where the Rayleigh-Jeans approximation consistently overestimates the shift in $L_{\nu}$ between frequency channels.  This suggests that the constant luminosity model results may indeed be biased, as expected.

To the authors' knowledge, no submillimeter debris disk survey has included such a large and unbiased sample of main sequence stars in spectral classes A-M as we have included here in our analysis. For this reason, a direct comparison of the baseline results presented in Fig.~\ref{fig:LogNorm Fixed Sigma Baseline Results} with constraints from the literature is not straight forward. Observations from the  {\it Herschel} DEBRIS (Disk Emission via a Bias-Free Reconnaissance in the Infrared/Submillimeter) survey, however, include stars with spectral types that are most closely matched to the sample used in this analysis. The DEBRIS survey observed a total of 446 stars in spectral classes A-M, and detected 77 debris disks for an incidence rate of $\sim 17\%$ \citep{10.1093/mnras/stx1378}. For reference, our constraint on $f_{\rm disk}$ for the fiducial star sample is $f_{\rm disk} = 10\% \pm^{7\%}_{4\%}$ at 68\% confidence. Indeed, our constraint on $f_{\rm disk}$ is consistent with measurements from the DEBRIS survey, but also allows for the possibility of a larger population of debris disks
at 95\% confidence. In \S\ref{sec: changing star selection}, we explore how our constraint on $f_{\rm disk}$ varies as a function of spectral type.


Our constraint on $\mu_{\ln{s}}$ is also reasonable, preferring slightly lower values than measurements with {\it Herschel} of the debris disk surrounding Fomalhaut. For reference, \citet{Herschel_Fomalhaut} measures an integrated flux over the Fomalhaut debris disk of 1.1~Jy, corresponding to a $\mu_{\ln{s}}$ of 93.3. In this section, we expect to find $\mu_{\ln{s}}$ preferring somewhat lower values than this measurement, given that Fomalhaut has a particularly bright debris disk. Indeed, we find this to be the case. In Appendix~\ref{app: individual stars}, we measure thermal emission from Fomalhaut's debris disk with {\it Planck} 857~GHz, and find constraints in agreement with measurements from {\it Herschel}.

\subsection{Debris disk constraints as a function of  stellar magnitude}
\label{sec: changing star selection}

\begin{figure}[t]
    \includegraphics[width=8.5cm]{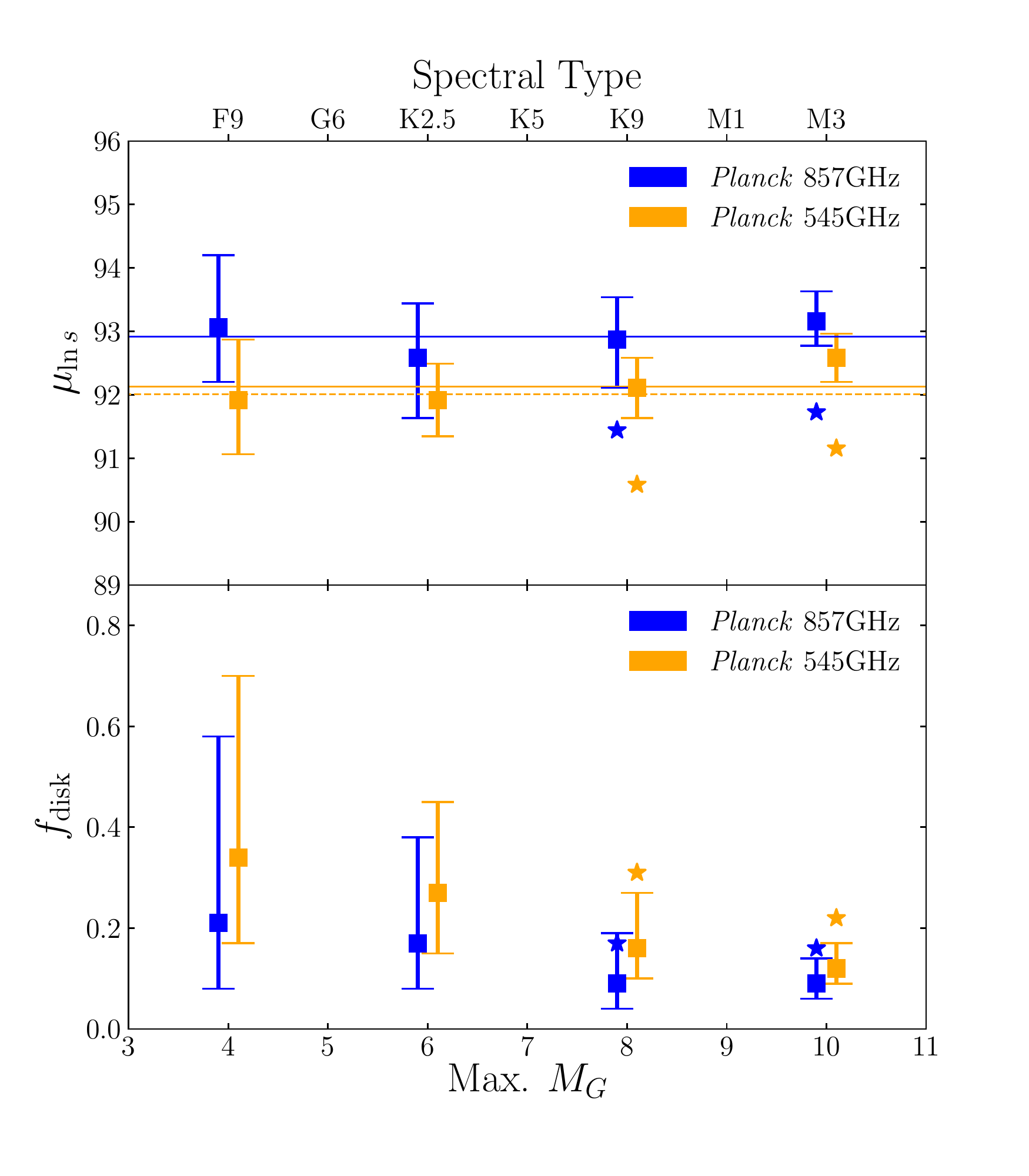}
    \centering
    \caption{Constraints on $\mu_{\ln s}$ and $f_{\rm disk}$ as a function of absolute magnitude cut, using the log-normal luminosity model with $\sigma_{\ln{s}} = 1.5$; error bars correspond to 68\% confidence intervals. 
    For Max. $M_G = [4,6,8,10]$, the number of stars included is $[311, 982,1872,3583]$. Including dim stars generally leads to lower values of $f_{\rm{disk}}$ with tighter constraints, the latter being a result of a rapid increase in the number of dim stars with large values of $M_G$. The orange and blue star markers in the final two bins indicate the best fit values when we fix $\sigma_{\ln{s}} = 2.0$ for the 545 and 857~GHz channels respectively. The dashed horizontal line in the top panel depicts the expected shift in $\mu_{\ln{s}}$ if the debris disk emission profile is well described by the Rayleigh-Jeans law, for which $L_{\nu} \propto \nu^2$. The location of the dashed orange line is scaled with respect to the average $\mu_{\ln{s}}$ in the $857$~GHz channel, shown as the solid blue line. The solid orange line is the average of the 545~GHz measurements, and is provided for reference. We generally find that our constraints on $\mu_{\ln{s}}$ are within $1\sigma$ for varying values of $M_G$, and the shift between frequency channels is consistent with the Rayleigh-Jeans assumption.}\label{fig:Vary Max M_G}
\end{figure}

We now consider how the debris disk constraints change as the stellar selection is varied.  Using the fiducial HI mask with 10\% of the sky left uncovered and a 80~pc maximum distance cut, we explore the dependence of the parameters $f_{\rm{disk}}$ and $\mu_{\ln{s}}$ on the maximum G band absolute magnitude of stars included in our sample. Results for this test are shown in Fig.~\ref{fig:Vary Max M_G}.  We  note that for all distance cuts explored on data with $M_G < 10$, the sample is most heavily dominated by spectral classes K-M  (see Appendix~\ref{app:spec_type} for more discussion).

Generally, we find that adding cool, dim stars to our analysis pushes the constraints on $f_{\rm{disk}}$ to lower values. In particular, for $M_G<4$, our stellar population is comprised of stars in spectral classes A-F. For this subset of stars, we find $f_{\rm disk} = 21\% \pm^{37\%}_{13\%}$. For $M_G < 8$, we measure spectral classes A-K, for which $f_{\rm disk} = 9\% \pm ^{10\%}_{5\%}$. For $M_G < 10$, we measure spectral classes A-M3, for which $f_{\rm disk} = 9\%\pm^{5\%}_{3\%}$. Uncertainties correspond to intervals of 68\% confidence, and are illustrated by the blue points in the bottom panel of Fig.~\ref{fig:Vary Max M_G}. We compare our constraints on $f_{\rm disk}$ to collective measurements from previous debris disk studies, taking care to only include the spectral types used in each ${\rm Max.} \ M_G$ bin. We generally find an agreement, with $\sim$ 24\%, 18\%, and 15\% of stars in spectral classes A-F, A-K, and A-M3 hosting disks, respectively \citep{10.1093/mnras/stu1864,Sibthorpe_2017,binks_2015}.



Constraints on $\mu_{\ln{s}}$ remain generally consistent, all within $1\sigma$, as the maximum $G$ band absolute magnitude is varied. The dashed horizontal line in the top panel of Fig.~\ref{fig:Vary Max M_G} depicts the expected shift in $\mu_{\ln{s}}$ under the Rayleigh-Jeans assumption, determined by averaging the data points in the 857~GHz channel and scaling appropriately by $(545/857)^2$. As we find in Fig.~\ref{fig:LogNorm Fixed Sigma Baseline Results}, the constraints from the 545~GHz channel are generally consistent with the expected frequency-dependent shift.

Interestingly, we do not find a significant transition to lower values of $\mu_{\ln{s}}$ as the number of dim stars in the sample increases.  This is somewhat surprising, as less luminous stars are expected to have cooler and smaller debris disks. One possible explanation for this finding is that  $\sigma_{\ln s}$ increases for fainter stars; as we have seen, this could bias the results to low $f_{\rm{disk}}$ and high $\mu_{\ln s}$.   For these reasons, it is possible that the the true $\mu_{\ln s}$ is lower than our results show in Fig.~\ref{fig:Vary Max M_G}, while the true $f_{\rm{disk}}$ is higher for the faint end of our sample. We illustrate this behavior by overplotting the best fit $\mu_{\ln{s}}$ and $f_{\rm{disk}}$ for $\sigma_{\ln{s}} = 2.0$ in Fig.~\ref{fig:Vary Max M_G}, for the two lowest luminosity bins  (our choice of $\sigma_{\ln s}=1.5$ is motivated by a somewhat brighter sample of stars, see Fig.~\ref{fig:75 Known Disks}). These measurements are represented by the orange and blue stars, corresponding to the best fits from the 545 and 857~GHz channels respectively.

\subsection{Candidate stars}
\label{sec:candidates}
Although our analysis is not aimed at detecting individual debris disk candidates, we can identify candidates using the individual likelihoods that we have computed.  We utilize the log-normal luminosity model with $\sigma_{\ln{s}} = 1.5$, as in the previous sections, to identify individual stars that contribute significantly to the total model likelihood. In particular, we use a likelihood ratio to compare two models for each star: in one model, $f_{\rm{disk}}$ is set to $1$ in Eq.~\ref{eq:likelihood_model}, with $\mu_{\ln{s}}$ set to the best fit value from the fiducial run on data. In the other model, $f_{\rm{disk}}$ is set to $0$ in Eq.~\ref{eq:likelihood_model}. We then define the log-likelihood difference between the two models:
\begin{multline*}
    \Delta \log P(d_i) \equiv \log P(d_i | f = 1, \mu_{\ln{s}} = \mu_{\ln{s},0}) \\
    - \log P(d_i | f = 0)
\end{multline*}
where $\mu_{\ln{s},0}$ is the best fit value from our fiducial result on data (\S\ref{sec:Log-Normal Results}). Stars with large $\Delta \log P$ are better described by a model with a debris disk emission signal than one without, so we rank all stars according to their respective values of $\Delta \log P$ and report the top contributors as possible debris disk candidates.

In Table~\ref{table:1}, we present the top 80 debris disk candidate stars with the highest $\Delta \log P$. Our choice of including 80 stars is not entirely arbitrary; on simulated data, we rank all mock stars, some with an added debris disk signal, by their respective $\Delta \log P$. We find that below a certain threshold value, the false positive rate begins to rise significantly. This threshold corresponds to a cutoff at approximately 80 stars on real data.

To construct Table~\ref{table:1}, we use the fiducial choice of a 10\% HI mask, but include stars out to 80~pc for a larger sample size. We utilize the highest resolution channel available at 857~GHz. 

Our approach for identifying candidates is sensitive to non-debris disk fluctuations in the {\it Planck} maps. These fluctuations tend to be easily differentiated from possible debris disk signals in the data by their extended profiles. Due to their radial extent and the size of the {\it Planck} beam, debris disks, on the other hand, are not expected to have a significant amount of flux in regions surrounding the central star-hosting pixel in an $N_{\rm side} = 1024$ \texttt{Healpix} map. We therefore place an asterisk in the ``Visual" column for possible debris disk signals that appear to satisfy this condition. We identify these stars as particularly strong candidates. We also include a column indicating whether or not a star is listed as a debris disk candidate in previous analyses. In order to determine the status of this column for each star, we use the publicly available SIMBAD astronomical database \citep{2000A&AS..143....9W} to mark stars with associated papers containing the key-word ``Debris Disk(s)" in the title. Stars possibly associated with debris disks under this criterion are marked with Y, while stars not satisfying this criterion are marked with N. The ordering of the stars in Table~\ref{table:1} is determined by the Rank \# of each star. This number specifies the ranking of each star's $\Delta \log P$ in the entire sample. A star with a Rank \# of 0 corresponds to the highest $\Delta \log P$, while subsequent Rank \#'s correspond to increasingly lower values of this metric.

Of the 80 stars with the top $\Delta \log P$, 24 are identified as previous debris disk candidates according to our criterion, while 43 are identified as particularly strong candidates from visual inspection.

\section{Discussion}
\label{sec:discussion}
\subsection{Summary}
We have presented constraints on the debris disk population around stars within roughly 100~pc of the sun, using a combination of {\it Planck} and {\it Gaia} data.  While the {\it Planck} survey has lower resolution and sensitivity compared to that of other surveys used for debris disk analyses, it is an all sky survey and thus enables comprehensive studies of a large sample of stars. Furthermore, the {\it Planck} wavelength coverage makes it well suited to studying the cold extended disks around dim stars. 

In our analysis, individual debris disks are not necessarily detected, so we use a likelihood approach that enables us to fit for the fraction of stars with disk-like emission, and for the disk luminosity. We apply different models for the disk luminosity distribution, and select a log-normal distribution with fixed width as our fiducial choice.

Our main result is shown in Fig.~\ref{fig:LogNorm Fixed Sigma Baseline Results}, and includes an 8.8$\sigma$ detection for the 80~pc distance cut in the 545~GHz frequency channel. The constraint on the fraction of main sequence stars that host debris disks, in a sample dominated by K and M-dwarf stars, is $f_{\rm disk} =  0.10 \pm^{0.165}_{0.07}$ at 95\% confidence. This constraint is consistent with analyses using {\it Herschel} and {\it Spitzer}.  \cite{Sibthorpe_2017,2013A&A...555A..11E,2008ApJ...674.1086T}, for instance, have measured hundreds of debris disks around a sample of typically brighter
stars using such surveys, and find $f_{\rm disk}$ in the range of 10-30\% .

We also place constraints on $f_{\rm disk}$ as a function of the absolute magnitude of the stars in our sample. Stars with large absolute magnitudes, i.e.  dim K and M stars, are especially interesting for present day studies of debris disks and planetary systems, since there are limited submillimeter observations of these systems \citep{2017MNRAS.469..579B}. In Fig.~\ref{fig:Vary Max M_G}, we find a preference for lower disk fractions as we allow fainter stars to enter our sample. At the 80~pc distance cut with $M_G < 10$, $\sim 75\%$ of the stars in our sample are in spectral classes K-M. For this set of stars, we find $f_{\rm disk} = 0.09 \pm^{0.11}_{0.05}$. It is not uncommon for analyses focusing on the detection of debris disks around M-dwarf stars to report detection rates as low as ours. \citet{2017MNRAS.469..579B}, for instance, report detection rates for young M-dwarf stars $\sim 13\%$, falling below $7\%$ for older M-dwarfs. Detection rates for K type stars, while slightly higher than that of M-dwarfs, are found to be  lower than the more luminous spectral classes \citep{Sibthorpe_2017}.

We test for thermal emission by comparing measurements in the 857 and 545 GHz channels with the expectation from the Rayleigh-Jean regime $\nu^2$ scaling. We find consistency with the Rayleigh-Jeans relation for the full sample of {\it Gaia} stars (see Fig.~\ref{fig:LogNorm Fixed Sigma Baseline Results} and Fig.~\ref{fig:Log-Normal 50 65 80pc Results with Sigma Prior}). For the dimmest stars in our samples, the flux in the 545 GHz channel shows hints of departures from the Rayleigh-Jeans scaling, consistent with colder debris disks. 

Since our analysis is carried out in submillimeter wavelengths, it is well suited to constraing the properties of cold/extended disks, especially around dim M-dwarf stars. In fact, there has been  mounting evidence for a substantial population of debris disks too cold to make the detection thresholds of shorter wavelength surveys such as WISE and IRAS \citep{Lestrade_2006,Lestrade_2009,2017MNRAS.470.3606H,2019arXiv191013142L}. Submillimeter debris disk surveys differ from their shorter wavelength counterparts, since they probe cooler disks and have stronger constraining power in the disk outskirts, which is expected to have larger dust grains and more massive planetesmials. At 857~GHz, our analysis is sensitive to such cool systems, and our results are suggestive of the possibility seen with {\it Herschel} \citep{Lestrade_2009}, that debris disk systems around dim stars could in fact be escaping detection by shorter wavelength surveys \citep{2019arXiv191013142L}. 

\subsection{Validation of our analysis and candidate debris disks}

While the main goal of our analysis is to provide constraints on the number of stars with debris disks, our approach is easily applied to a small subset of stars, or even measurements around individual stars. In Fig.~\ref{fig:75 Known Disks}, we verify that our analysis, applied to a set of well studied debris disks, yields measurements on $f_{\rm{disk}}$ consistent with 100\%.  In Fig.~\ref{fig:Beta_Pictoris_Fomalhaut}, we show the agreement of our flux measurements for two well studied debris disks to measurements from {\it Herschel}. 

We also identify like debris disk candidates by ranking each star according to its likelihood of hosting a debris disk. See Table~\ref{table:1} for the 80 most promising candidate stars, most of which have not been identified as debris disk candidates previously.

\subsection{Caveats}

A potential source of bias in our measurements could stem from the luminosity distribution model we adopt. Our fiducial analysis relies on modeling the debris  luminosity  distribution as log-normal, with a fixed width based on our analysis of a population of known debris disk (\S\ref{sec:Log Normal Model on Known Disks}). We choose to keep the width of the log-normal  distribution fixed in order to avoid possibly fitting background fluctuations in the maps.  With higher  resolution and higher signal-to-noise data from future surveys, we expect to be able to treat $\sigma_{\ln s}$ as a free parameter.

Another caveat for our analysis concerns our extensive masking of the sky.  By excluding regions of  the sky with high  HI  column densities, we ensure that our analysis is more robust to contamination from galactic backgrounds.  However, it is possible that this masking leads to a selection of stars whose debris disk properties differ from the global average.  For instance, if debris disks are more likely in star forming regions, and the HI mask preferentially excludes these regions, we could bias the constraints on $f_{\rm disk}$ low relative to the true, all-sky value.  Our  constraints should therefore be viewed as applying only to stars in the unmasked region of our analysis. Additionally, foreground extinction could potentially impact the measured signal, though this scenario is unlikely given that we do not include stars past 80~pc in our sample.

While our analysis is meant to provide constraints on a population of debris disks, our approach is potentially sensitive to other sources of thermal emission.  Three possibilities are:  protoplanetary disks,   nebular  emission  seen  in  young  stars, or a  “halo”  of  particles  ejected  from a disk  by  radiation  pressure  or  stellar  winds (see \citet{2015Ap&SS.357..103W,2015A&A...573A...6P,2017ApJ...848....4C}, for example).   The  ages  of  the  hot stars in our sample are typically larger than 100 Myr, so the former two explanations appear unlikely.  Distinguishing a particle “halo” appears challenging, as the emission in both cases is dominated by small grains. Interestingly, in Table 1 our list of 80 candidate sources includes one flare star.

\subsection{Prospects with  CMB datasets}

The present analysis is in some sense a proof of principle; we have demonstrated that one can use wide-field CMB-focused data to constrain debris disks. It is worth considering how future CMB observations could be used to place tighter constraints on debris disks, and what we could hope to learn from such studies in the future. 

Measurements from {\it Planck} are hindered by a large beam size relative to the angular scale of the typical debris disk. Higher angular resolution data would be beneficial to an analysis like ours, by reducing the impact of background sources contaminating the debris disk signals. 
Galactic dust in particular  has less power on very small scales. 

A central concern of this analysis is placing constraints on a population of debris disk systems surrounding cool K-M type stars. Such systems are poorly understood, and have received an increasing amount of attention as strategic targets for future surveys \citep{2019arXiv190602129D}. There are several current and future CMB datasets well suited to placing tighter constraints on these systems. Ongoing surveys include  SPT-3G \citep{2014SPIE.9153E..1PB} and Advanced ACTPol \citep{2016JLTP..184..772H}, while the planned surveys with the Simons Observatory \citep{2019JCAP...02..056A} and CMB-S4 \citep{2016arXiv161002743A} will provide significantly higher resolution and sensitivity in the submillimeter sky compared with the {\it Planck} maps used in this analysis. These surveys are lacking in one respect relative to {\it Planck}, namely its high frequency channels. 

Like the analysis carried out by \citet{Baxter:2018} on measuring thermal emission from exo-Oort clouds with {\it Planck}, one of the biggest limitations to our analysis is the modeling of backgrounds. With higher resolution data, background modeling could be significantly improved, allowing more stars to enter our sample yielding better statistics. Additionally, higher resolution data could potentially lead to resolved observations of nearby debris disks, adding to our current understanding of finer scale disk structure without carrying out pointed measurements.

Finally, wide area and high resolution CMB data, combined with a likelihood analysis such as our own, could be used to constrain other poorly understood systems such as debris disks around white dwarfs \citep{2012ApJ...760...26B}, transient emission events from flare stars \citep{MacGregor_2018} or extragalactic sources. Such sources of emission, while not the focus of this analysis, are well suited to detection by the methods developed in this paper. 

\section*{Acknowledgments}
We thank  Mark Devlin, Mike Jarvis, Greg Laughlin and Nicole Pawellek for stimulating  discussions related to this work. We are grateful to Cullen Blake for numerous discussions and helpful comments on the paper. 

\begin{table*}
\centering                                  
\begin{tabular}{c c c c c c}          
\hline\hline                        
Rank \# & Name/ID & Type & Dist. [pc] & Visual & Prev. Candidate \\    
\hline                                   
    0  & 36 UMa & F8V C & 12.9 & * & Y \\      
    1  &  CD-49 424  &  K5-V C  &  39.4  &  ...  & Y  \\
    2  &  LP 221-55  &  M0.5Ve   &  41.1   &  * & N \\
    3  &  TYC 7501-1011-1  &  M1 D  &  33.6  &  ...  & N   \\
    4  &  CD-53 694  &  ...  &  51.0  &  *  & N \\
    5  &  Ross 917  &  M0.5Ve C  &  35.7  &  ...  & N \\
    6  &  L 177-19  &  ...  &   35.9  &  ...  & N \\
    7  &  BD+26 2415  &  K5 D  &  47.2  &  *  & N \\
    8  &  HD 87141  &  F5V D  &  51.9  &  ...  & Y \\
    9  &  CD-28 1030 - Flare Star  &  M1Ve  &  19.6  &  *  & N \\
    10  &  HD 120004  &  F8 E  &  55.7  &  *  & Y \\
    11  &  phi Gru  &  F4V C  &  34.1  &  *  & Y \\
    12  &  G 267-3  &  ...  &  32.4  &  *  & N \\
    13  &  CD-47 277  &  M0V(e) C  &  42.0  &  ...  & N \\
    14  &  CD-27 470  &  ...  &  76.0  &  *  & N \\
    15  &  G 56-19  &  ...  &  40.7  &  *  & N \\
    16  &  BD+62 1259  &  G5 D ~  &  65.3  &  *  & N \\
    17  &  HD 129499  &  G5 E  &  53.9  &  *  & Y \\
    18  &  LP 768-488  &  ...  &  51.4  &  *  & N \\
    19  &  NLTT 1733  &  ...  &  50.6  &  ...  & N \\
    20  &  HD 8406  &  G3V C  &  37.7  &  ...  & Y \\
    21  &  Gaia DR2 2310240437249834496  &  ...  &  35.2  &  ...  & N \\
    22  &  TYC 3034-841-1  &  ...  &  68.1  &  *  & N \\
    23  &  UCAC2 11235167  &  ...  &  26.7  &  *  & N \\
    24  &  HD 11112  &  G3V C  &  44.4  &  *  & Y \\
    25  &  UCAC4 526-056995  &  ...  &  45.5  &  *  & N \\
    26  &  G 60-2  &  K7 D  &  59.0  &  ...  & N \\
    27  &  BD+41 2695  &  K5V C   &  15.8  &  ...  & N \\
    28  &  UCAC4 784-023772  &  ...  &   57.0  &  ...  & N \\
    29  &  UCAC4 514-055849  &  M2.0V C  &  26.6  &  *  & N \\
    30  &  V* BK CrB  &  K7V C   &  52.8  &  ...  & N \\
    31  &  LP 888-47  &  ...  &   68.6  &  *  & N \\
    32  &  LP 99-246  &  M3 D  &  37.5  &  *  & N \\
    33  &  HD 219122  &  G1/2V D  &  77.4  &  *  & Y \\
    34  &  NLTT 32168  &  ...  &  69.9  &  *  & N \\
    35  &  HD 19545  &  A3V C  &  54.2  &  *  & Y \\
    36  &  35 LMi  &  F3V D  &  46.3  &  *  & Y \\
    37  &  HD 91312  &  A7IV C ~  &  33.5  &  *  & Y \\
    38  &  LTT 9738  &  ...  &  60.3  &  ...  & N \\
    39  &  2MASS J03013462-5743097  &  ...  &  56.7  &  *  & N \\
    40  &  L 175-9  &  ...  &  35.6  &  ...  & N \\
    41  &  HD 211369  &  K2.5Vk: C  &  25.9  &  ...  & Y \\
    42  &  StKM 1-902  &  M1.0Ve C  &  41.3  &  ...  & N \\
    43  &  BD+58 1323   &  K2V C  &  39.0  &  *  & N \\
    44  &  HD 11608  &  K1/2V D  &  43.2  &  *  & Y \\
    45  &  2MASS J04593244-2924122  &  ...  &  30.8  &  ...  & N \\
    46  &  WISEA J010746.74-525505.9 &  ...  &  79.5  &  ...  & N \\
    47  &  BD+47 2312  &  G5 D ~  &  39.6  &  ...  & Y \\
    48  &  HD 223171  &  G2V C  &  41.4  &  ...  & Y \\
    \hline    
\end{tabular}
\end{table*}

\begin{table*}[ht!]
\vspace{-7.5cm}%
\centering                                  
\begin{tabular}{c c c c c c}          
\hline\hline                        
Rank \# & Name/ID & Type & Dist. [pc] & Visual & Prev. Candidate \\    
\hline                                   
    49  &  HD 151044  &  F8V C  &  29.3  &  *  & Y \\
    50  &  HD 116956  &  G9V E ~  &  21.7  &  *  & Y \\
    51  &  G 146-18  &  ...  &  40.7  &  ...  & N \\
    52  &  V* EK Dra   &  G1.5V E ~  &  34.4  &  *  & Y \\
    53  &  HD 149026  &  G0IV C  &  76.0  &  *  & Y \\
    54  &  V* AT CrB  &  K3V C  &  23.2  &  *  & Y \\
    55  &  UCAC4 730-050199  &  ...  &  37.7  &  ...  & N \\
    56  &  HD 111631  &  M0V C  &  10.7  &  *  & Y \\
    57  &  UPM J2145-4145  &  ...  &  76.6  &  ...  & N \\
    58  &  G 116-59  &  ...  &  50.4  &  ...  & N \\
    59  &  StKM 1-970  &  K5 D  &  44.2  &  ...  & N \\
    60  &  LP 1032-111  &  K4 D  &  58.7  &  *  & N \\
    61  &  CD-28 1222  &  ...  &  59.6  &  ...  & N \\
    62  &  BD+40 2790  &  K0 D ~  &  42.0  &  ...  & N \\
    63  &  HD 95175  &  K3 D  &  32.2  &  ...  & N \\
    64  &  G 178-24  &  ...  &  79.9  &  *  & N \\
    65  &  HD 238600   &  K0 E  &  40.0  &  ...  & N \\
    66  &  HD 108944  &  F9V C  &  50.0  &  *  & Y \\
    67  &  TYC 8082-471-1  &  ...  &  43.2  &  ...  & N \\
    68  &  LEHPM 6585  &  ...  &  66.1  &  *  & N \\
    69  &  2MASS J23233534-4451522  &  ...  &  50.0  &  *  & N \\
    70  &  BD-14 363  &  G5V D ~  &  64.8  &  ...  & N \\
    71  &  2MASS J01182472-2611189  &  ...  &  71.6  &  *  & N \\
    72  &  G 148-61  & M1.5Ve C &  25.9  &  *  & N \\
    73  &  LP 379-98  &  M0V C  &  44.8  &  ...  & N \\
    74  &  V* DQ Tuc  &  M2V C  &  29.1  &  ...  & N \\
    75  &  TYC 8822-1125-1  &  ...  &  63.7  &  ...  & N \\
    76  &  HD 212038  &  K2.5V C   &  27.0  &  ...  & Y \\
    77  &  LSPM J1055+2455  &  ...  &  68.4  &  *  & N \\
    78  &  G 122-8  &  M3.5Ve C  &  23.2  &  ...  & N \\
    79 & WISEA J121643.04+322830.9 & ... & 69.1 & * & N
    \\
    \hline

\end{tabular}
\caption{Debris disk candidate stars ranked by their probability of hosting a debris disk. Stars with lower rank numbers have a greater $\Delta \log{P}$, and are thus considered better candidates. Stars which pass a visual inspection in the data are identified as especially promising candidates, marked by *. The Prev. Candidate column indicates whether or not the star of interest has been included in previous debris disk analyses according to the Simbad astronomical database. }
\label{table:1}
\end{table*}

\clearpage
\bibliography{thebib}

\appendix

\section{Asymptotic behavior of signal-to-noise with increasing distance cuts}
\label{app:snr}

We now consider how the constraints on disk properties are expected to behave as we increase the maximum stellar  distance cut.  For this purpose, we can assume that every star hosts a debris disk. As above, we refer to the data for the $i$th star as $d_i$, which is at a  distance $r_i$.  We additionally assume that the (scaled) luminosity is a constant, $s$, and that the noise  variance  is the same  for all  stars,  $\sigma_N^2$.  

The log-likelihood for the data is then
\begin{eqnarray}
L(d_i|s) &=& -\frac{1}{2} \sum_i \frac{(d_i - s/r_i^2)^2}{\sigma_N^2},
\end{eqnarray}
and the Fisher matrix for the single  parameter $s$ is
\begin{eqnarray}
F_{ss} &=& -\frac{\partial^2}{\partial s^2} \ln L \\
&=& \sum_i \frac{1}{\sigma_N^2 r_i^4}.
\end{eqnarray}
Let us assume that  the distribution of stars in the volume is such that the number density goes as $n(r) \propto r^{\alpha}$.   For uniformly distributed stars, we  would have $\alpha = 0$.  We then have  that  the  probability of  a star being in the radial bin between $r$  and $r+dr$ is $P(r) dr \propto r^{\alpha + 2} dr$. We can then approximate the Fisher matrix as
\begin{eqnarray}
F_{ss} &\sim& \int dr \,  r^{2+\alpha} \frac{1}{\sigma_N^2 r^4} \\
&=& \frac{\left[ r_{\rm max}^{\alpha - 1} -  r_{\rm min}^{\alpha - 1} \right]}{\sigma_N^2 (\alpha - 1) }.
\end{eqnarray}
Consequently,  the error  on $s$ goes like
\begin{eqnarray}
\sigma_s = \sqrt{F^{-1}_{ss}} \sim \frac{\sigma_N \sqrt{1-\alpha}}{\sqrt{r_{\rm min}^{\alpha - 1} -  r_{\rm max}^{\alpha - 1}}}.
\end{eqnarray}
For uniformly distributed stars with $\alpha = 0$, we have 
\begin{eqnarray}
\sigma_s \sim \frac{\sigma_N }{\sqrt{r_{\rm min}^{- 1} -  r_{\rm max}^{ - 1}}}.
\end{eqnarray}
In the limit that $r_{\rm max} \rightarrow \infty$, the error approaches a constant value, namely
\begin{eqnarray}
\sigma_s(r_{\rm max} \rightarrow \infty) \sim \sigma_N \sqrt{r_{\rm min}}.
\end{eqnarray}
Consequently, under the constant  luminosity model, we do not expect the constraints to  get arbitrarily tight as we  include  more and more distant stars.

\section{Additional simulation results}
\label{app: simulations}
In this appendix we consider further results of our analysis applied to simulated data with both Gaussian white noise backgrounds and the {\it Planck} backgrounds themselves.

\subsection{Gaussian simulation results}

\begin{figure}[h!]
    \includegraphics[width=8.5cm]{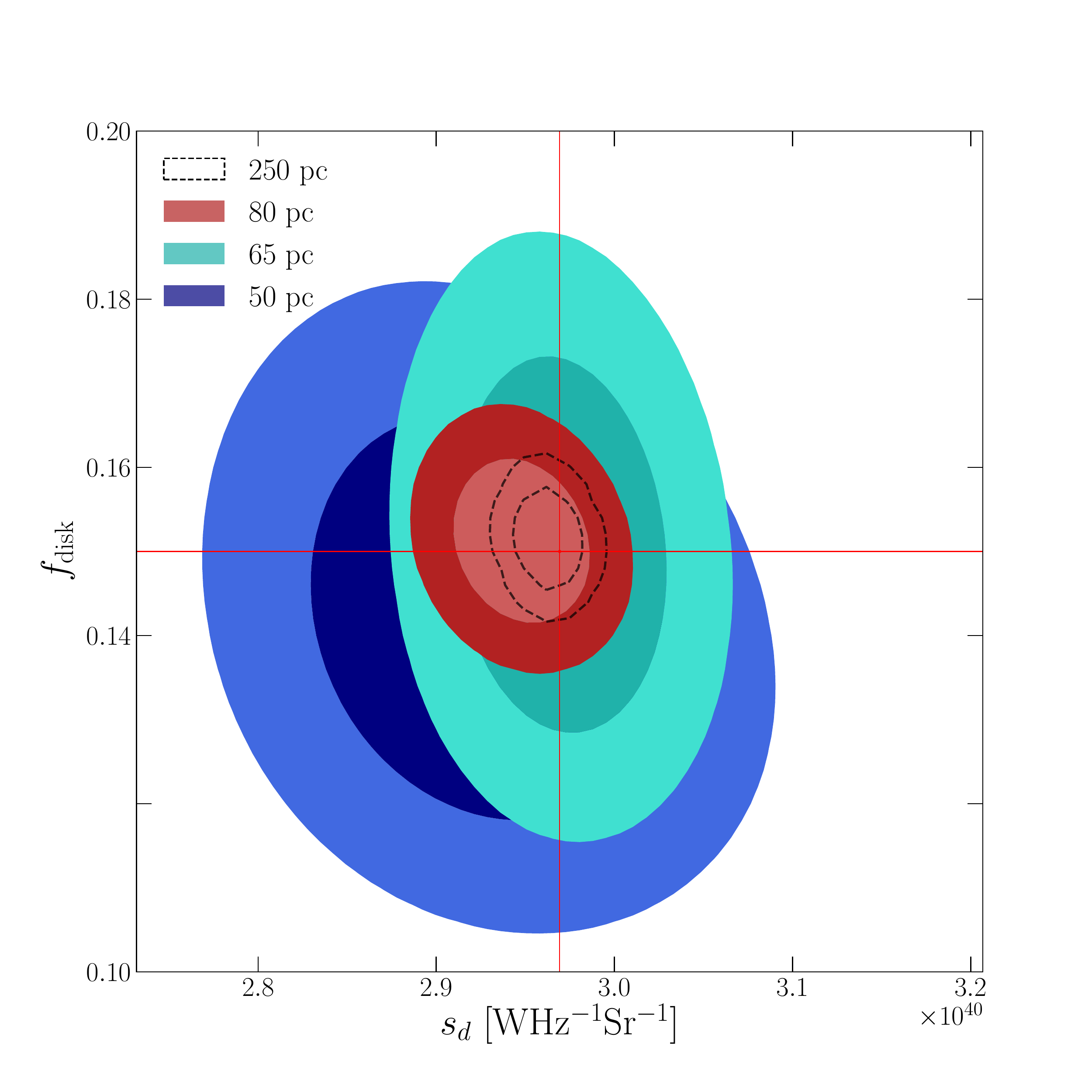}
    \centering
    \caption{Parameter constraints recovered from simulations with the Gaussian backgrounds for varying maximum distance cuts. Contours depict regions of 68\% and 95\% confidence, where the red crosshair depicts the input parameters. Each realization uses 5 times the number of stars found in real data. The model recovers the input disk fraction, $f_{\rm disk}$, and the signal amplitude, $s_d$, well within the 68\% region for each distance cut. 
    }\label{fig:Gaussian Simulation}
\end{figure}

In Fig.~\ref{fig:Gaussian Simulation} we present the results of our analysis applied to the Gaussian noise simulations,  as a function of maximum distance cut.  For each distance cut, we use five times the number of stars found in real data, after imposing the masking described in \S\ref{sec:mask} and the stellar selection cuts outlined in \S\ref{sec: Star Selection}. This corresponds to approximately 5000, 9800, 17300, and 5$\times10^5$ stars for the 50, 65, 80, and 250~pc distance cuts respectively.  We find that in all cases, the input parameters are recovered without bias.  As the maximum distance is increased, the constraints get tighter.   The constraints do not tighten as $1/\sqrt{N}$, where $N$ is the number of stars, since more distant stars have weaker signal.

\subsection{Potential for bias with the constant luminosity model}
\label{app: constant luminosity bias}
The constant luminosity model is prone to potentially biased results if the true underlying debris disk population is not well approximated by a constant luminosity model.  We test the extent of this bias by applying the constant luminosity model to a sample of mock debris disks with luminosities drawn from a log-normal distribution. 

\begin{figure}[t]
    \includegraphics[width=8.5cm]{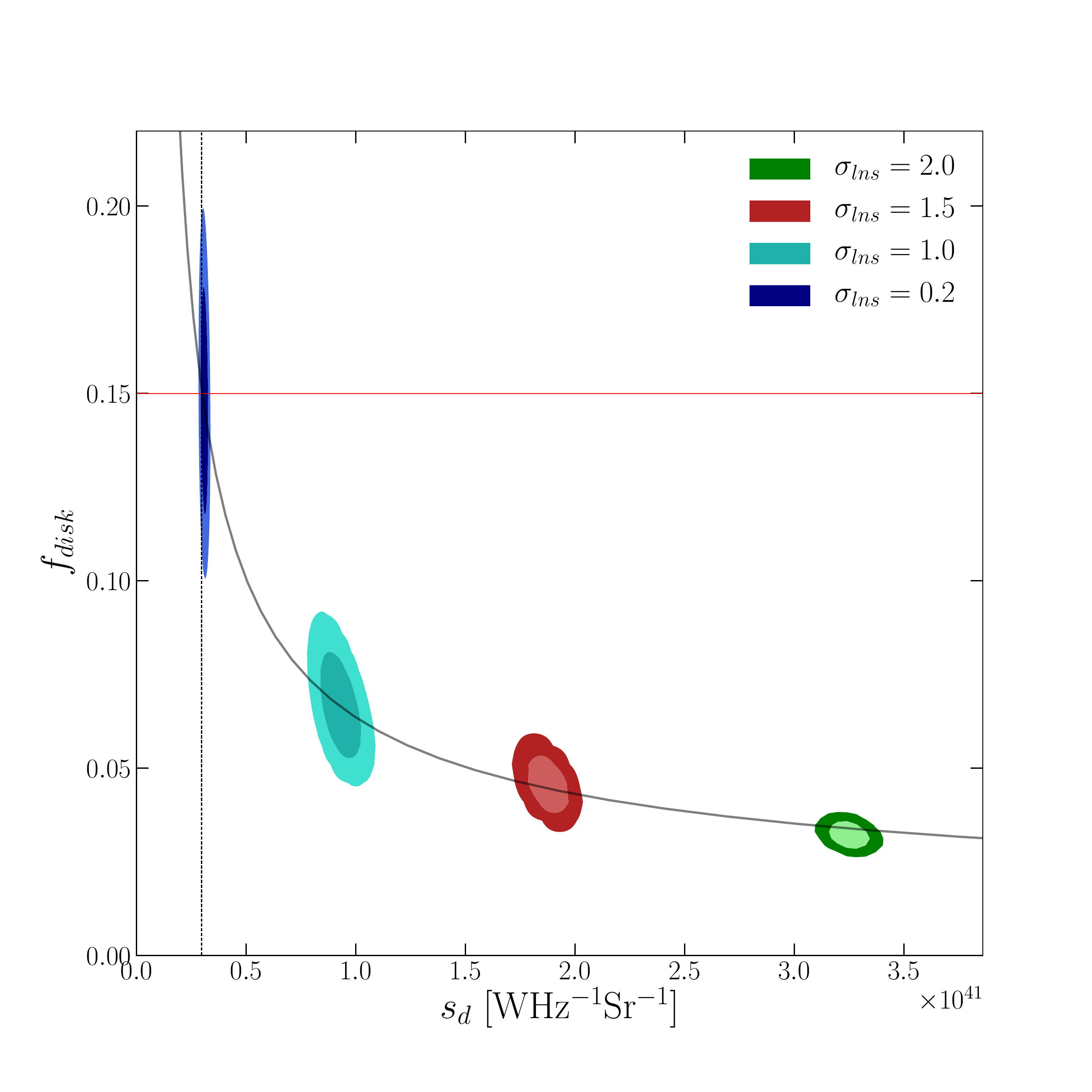}
    \centering
    \caption{Constraints recovered from the {\it Planck} background simulations using the constant luminosity model, where the true mock debris disk luminosity distribution is log-normal. Here, $\sigma_{\ln{s}}$ refers to the width of the underlying luminosity distribution, as in Eq.~\ref{eq:lognormal}. The input $f_{\rm disk}$ is shown as the horizontal red line, while the underlying mean of the luminosity distribution is illustrated as the dotted vertical line. We find increasingly biased results for $\sigma_{\ln{s}} \gtrapprox 0.5$, where luminous outliers dominate the model likelihood. We illustrate the degeneracy of the model parameters by over plotting a curve of constant total flux, effectively the relation $s_d f_{\rm{disk}} = \rm{constant}$.}\label{fig:ConstLum_withLogNorm}
\end{figure}

In Fig.~\ref{fig:ConstLum_withLogNorm}, we present results for this test with four different values of $\sigma_{\ln{s}}$. The input parameters are shown as dotted and solid red lines. We find that in the limit of small $\sigma_{\ln{s}}$, the constant luminosity model is sufficient in providing a minimally biased description of the debris disk population. For large $\sigma_{\ln{s}}$, however, luminous disks at the tail end of the log-normal distribution tend to drive $s_d$ high and $f_{\rm{disk}}$ low, consistent with the relation for constant total flux: $s_d f_{\rm{disk}} = \rm{constant}$. We overplot this degenerate relation in Fig.~\ref{fig:ConstLum_withLogNorm} as the solid black curve.

\begin{figure}[tp!]
    \includegraphics[width=8.5cm]{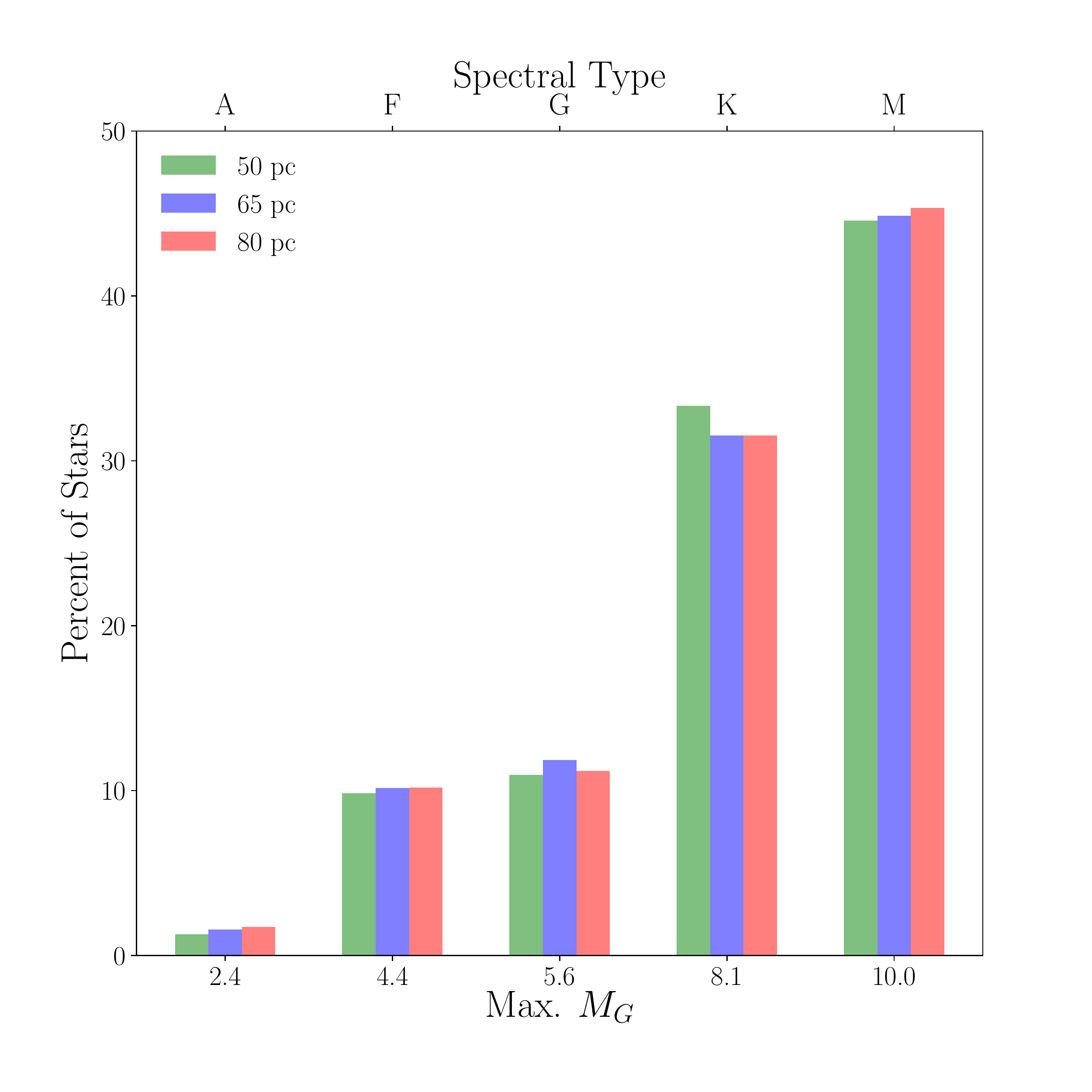}
    \centering
    \caption{The percent of stars by spectral type in our sample, after imposing the cuts described in \S\ref{sec: Star Selection} and the conservative 10\% HI mask. For each distance cut our sample is most heavily dominated by K-M stars.}\label{fig:Spectral Types}
\end{figure}

\begin{figure*}[h!]
    \includegraphics[width=18.2cm]{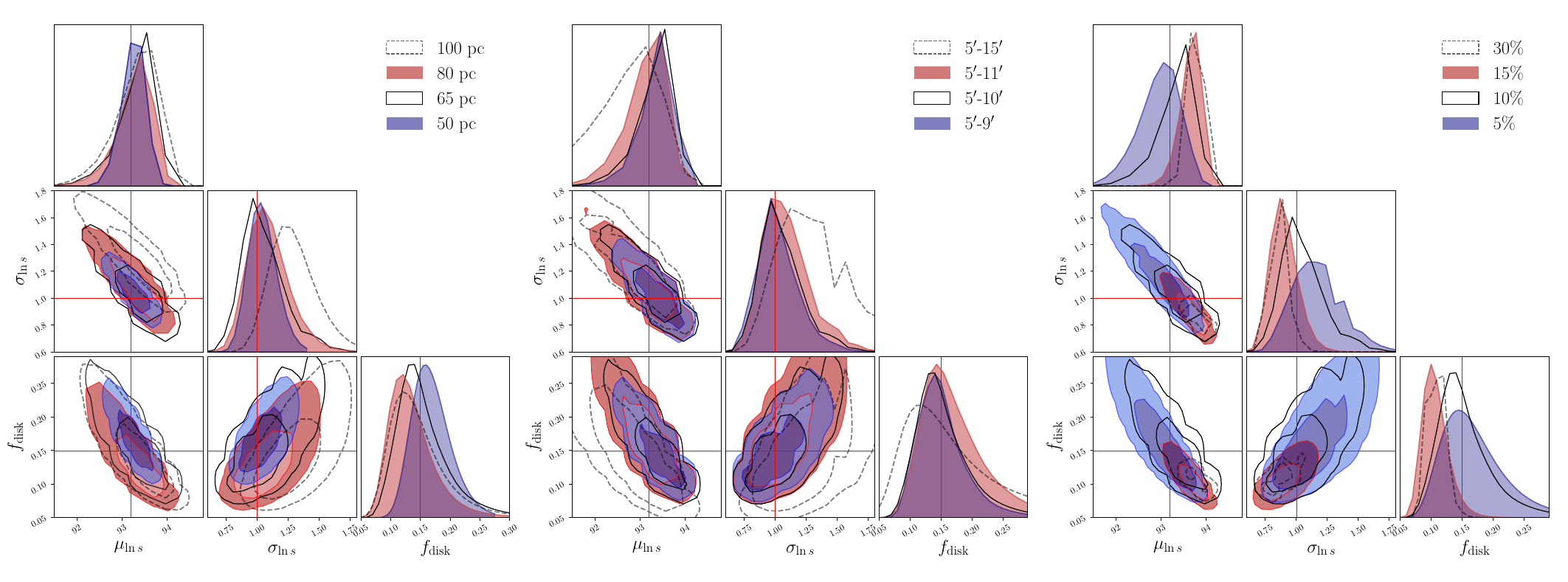}
    \centering
    \caption{{\it Planck} background simulation results for the log-normal luminosity model. As in Fig.~\ref{fig:Constant Luminosity Three Panel}, we use five times the number of stars found in real data for each realization. \textbf{Left:} Results for varying distance cuts with a fixed HI mask of 10\% and annulus width of 5 arcmin. \textbf{Middle:} Results for varying annulus width. For each realization, $\theta_{\rm{min}} = 5 \ \rm{arcmin}$. \textbf{Right:} Results for varying HI masks, where the percent value refers to the percentage of the sky used in the simulation. We use these simulation results to motivate our fiducial analysis choices.}\label{fig:Planck Simulation Lognormal}
\end{figure*}

\begin{figure*}[h]
    \includegraphics[width=17.5cm]{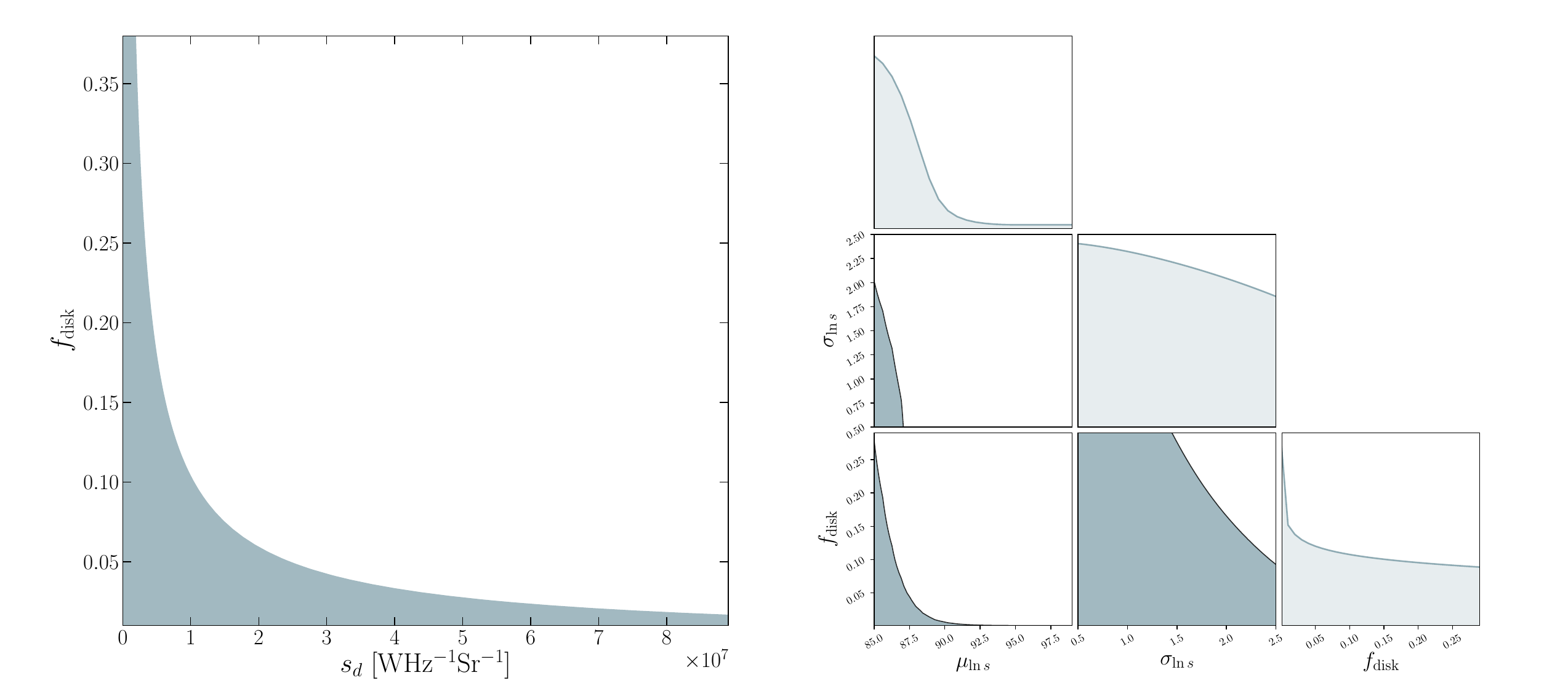}
    \centering
    \caption{Null test results for both the constant luminosity model (left) and the log-normal luminosity model (right). Both panels show results consistent with no signal, i.e. $f_{\rm{disk}}= 0$. Null tests are performed by choosing $N$ pixels in the {\it Planck} map that do not contain {\it Gaia} stars. Here, we also impose the 10\% HI mask discussed in \S\ref{sec:planck_sim_results} and resample distances from real {\it Gaia} stars. The analysis is then centered on these pixels to verify that a significant signal above background levels is not measured. We choose $N = 3000$ pixels. 
    }\label{fig:Null Test}
\end{figure*}

\subsection{Log-Normal luminosity model simulation results}

In Fig.~\ref{fig:Planck Simulation Lognormal}, we present the full set of simulation results for the log-normal luminosity model with mock debris disk signals added to the {\it Planck} backgrounds. In this case, the signal amplitudes are drawn from a log-normal distribution. A detailed description of the simulation procedures and discussion of these results can be found in $\S\ref{sec:planck_sim_results}$.  

We find that our analysis recovers the input parameters to within the errors.  Furthermore, for reasonable variations around our fiducial analysis choices, the constraints are minimally impacted.

\section{Further Results From Data}
\label{app: Results with Data}

\subsection{Spectral Types Populating our Sample}
\label{app:spec_type}
In Fig.~\ref{fig:Spectral Types}, we plot the percentage of main sequence stars in each spectral type for the three distance cuts used on data. Main sequence stars are selected by imposing the cuts outlined in \S\ref{sec:mask}, and the spectral type is determined using the provided {\it Gaia} G band magnitude for each star.  We find that for each distance cut, K-M stars make up the majority of our sample for $M_G < 10$. 

\subsection{Null Test Results}
\label{app: Null Test Results}
As described in \S\ref{sec:null_test}, we perform a null test by applying our analysis to pixels in the {\it Planck} map that are not hosts to {\it Gaia}-detected stars. In this case, our analysis should recover constraints consistent with $f_{\rm disk} = 0$.  We use $3000$ pixels for each null test, since this is approximately the number of stars included in our fiducial star sample. We assign each pixel used in the null test a distance drawn from the true distribution of {\it Gaia} stellar distances. The fiducial distance cut of 65~pc is used, with an HI mask leaving 10\% of the sky uncovered.  

In Fig.~\ref{fig:Null Test} we show the results of the null tests; we find constraints consistent with no signal for both the constant luminosity model (left panel) and the log-normal luminosity model (right panel). For the log-normal luminosity model null test, we impose the same prior on $\sigma_{\ln{s}}$ as we impose when centering the analysis on {\it Gaia} stars; namely, $0.5<\sigma_{\ln{s}}<2.5$. We discuss the motivation for this prior in Appendix~\ref{lognorm appendix}.

\subsection{Constant luminosity model results}
\label{app:Constant Lum}

\begin{figure}[t]
    \includegraphics[width=8.5cm]{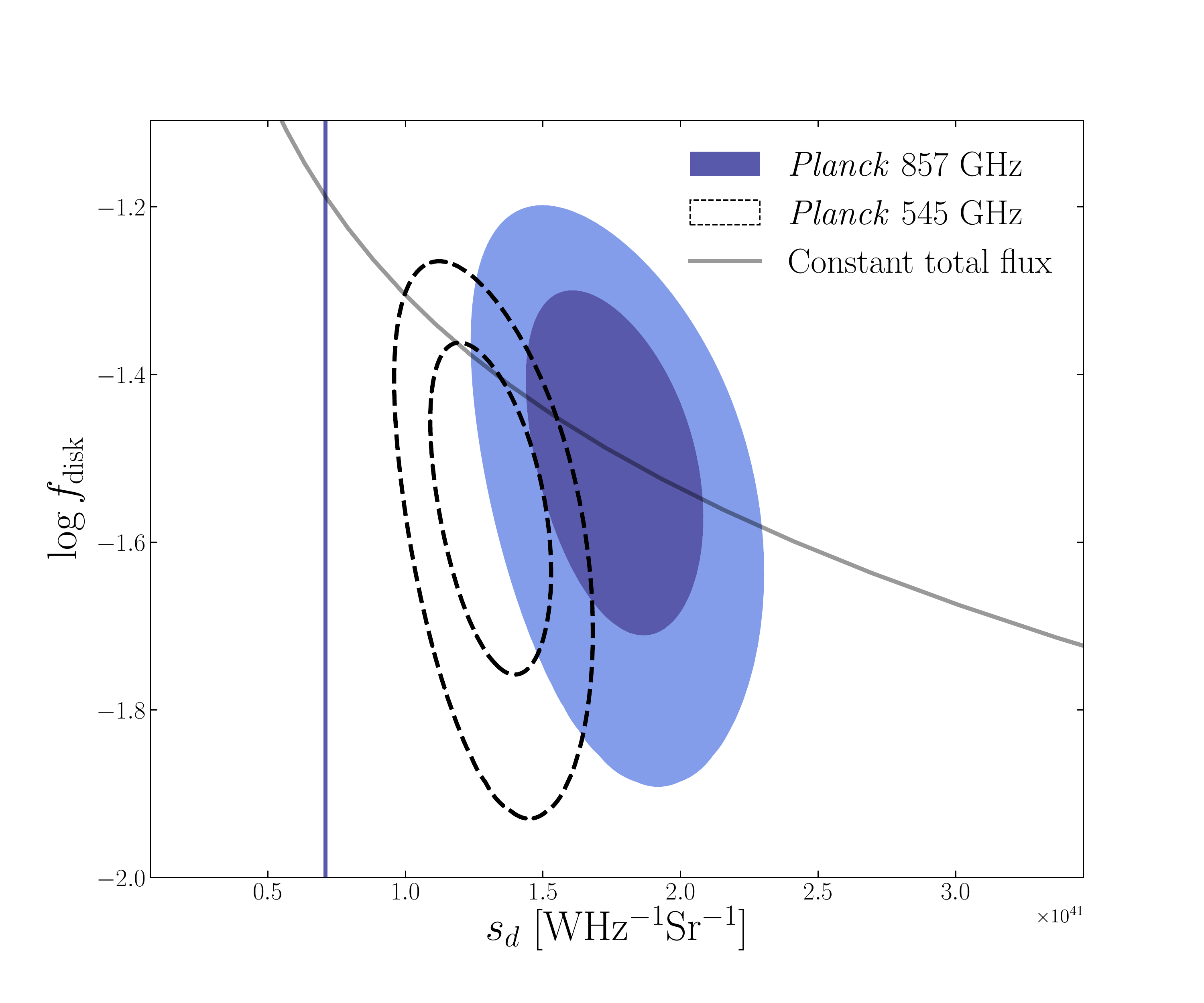}
    \centering
    \caption{Constant luminosity model results on real data from the {\it Planck} 857 and 545~GHz channels. The fiducial set of analysis choices is used, with a maximum distance cut of 65~pc, an annulus size of $\theta_{\rm{min}}=5$, $\theta_{\rm{max}}=10$ arcmin, and an HI mask leaving 10\% of the sky uncovered. The number of {\it Gaia} stars left is 1959. The solid blue and dashed contours represent parameter constraints derived from the 857 and 545~GHz {\it Planck} maps respectively. The vertical line depicts the expected shift in $s_{d}$ for an unbiased model, if the disk population is well approximated by the Rayleigh-Jean's law. We overplot the same degeneracy profile shown in Fig.~\ref{fig:ConstLum_withLogNorm}.}\label{fig:Const_Lum_Results}
\end{figure}

In Fig.~\ref{fig:Const_Lum_Results}, we present results on {\it Gaia} stars in the {\it Planck} 545 and 857~GHz channels using the constant luminosity model developed in \S\ref{sec:modeling}.  This figure adopts our fiducial analysis choices, with a 65~pc distance cut and an HI mask leaving 10\% of the sky uncovered. The remaining stars are selected according to the constraints outlined in \S\ref{sec: Star Selection}. Solid blue and dashed contours correspond to constraints derived from the 857 and 545~GHz maps respectively.   We  remind the reader that we have reason to believe that the parameter constraints inferred when assuming the constant luminosity model are biased.

The vertical line represents the expected shift in spectral luminosity from the Rayleigh-Jeans law ($I_\nu \propto \nu^2$) for an unbiased model, in going from the 857~GHz channel to the 545~GHz channel. We expect that the majority of stars in our sample obey such a relation.

We note that the constraints on $s_d$ with the constant luminosity model tend to slightly larger values than known flux measurements of Fomalhaut. Meanwhile, the constraints on $f_{\rm disk}$ prefer significantly lower values than what is commonly reported in the literature, usually in the range of $15-25\%$ for main sequence stars \citep{2014prpl.conf..521M}. 
As suggested by our tests on simulated data (and illustrated by Fig.~\ref{fig:ConstLum_withLogNorm}), the preference for low values of $f_{\rm disk}$ may be driven by the fact that the constant luminosity model is not a good description of the underlying debris disk population.

\subsection{Log-Normal luminosity model results with varying $\sigma_{\ln s}$}\label{lognorm appendix}
We now consider results from the {\it Planck} 545 and 857~GHz channels when applying the log-normal analysis with varying $\sigma_{\ln s}$. The stellar population used in this section is the same as in \S\ref{sec:Log-Normal Results}. 

\begin{figure}[t]
    \includegraphics[width=8.5cm]{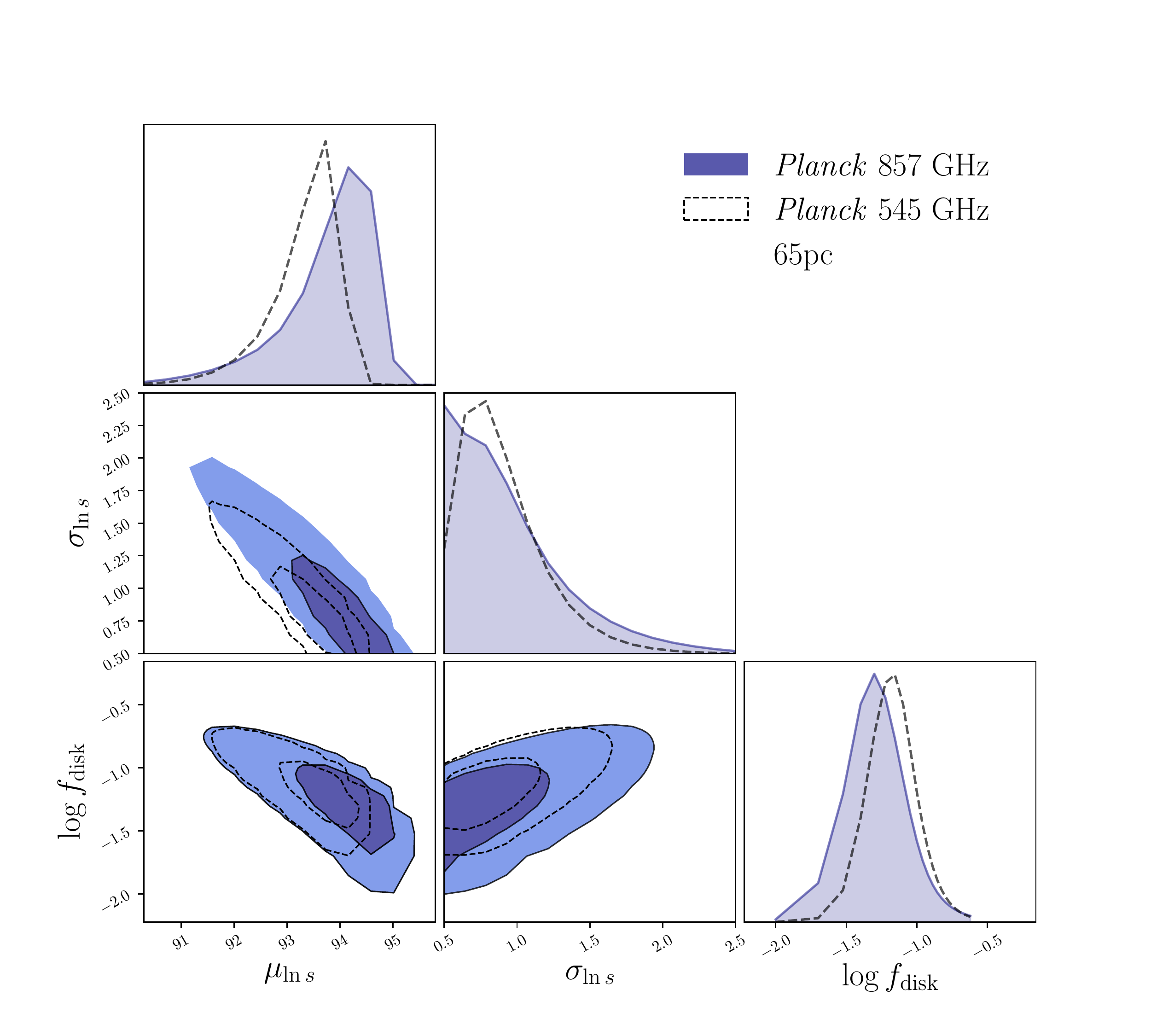}
    \centering
    \caption{Log-normal luminosity model results from the {\it Planck} 545 and 857~GHz channels, with a maximum distance cut of 65. Here, $\sigma_{\ln s}$ is varied. We use an HI mask leaving 10\% of the sky uncovered, and impose the restriction $M_G < 10$. The total number of stars is $\sim$ 2000.}\label{fig:Log-Normal 50 65 80pc Results with Sigma Prior}
\end{figure}

As described in \S\ref{sec:Log-Normal Results}, we expect our measurements to have fairly low signal-to-noise. Consequently, we impose a prior on $\sigma_{\ln{s}}$ of $\sigma_{\ln s} \in [0.5,2.5]$.  Results for the log-normal model analysis are shown in Fig.~\ref{fig:Log-Normal 50 65 80pc Results with Sigma Prior}. Marginalizing over $\sigma_{\ln{s}}$, we find constraints
\begin{equation*}
    \begin{split}
    \mu_{\ln{s}} & = 94.2 \pm^{1.29}_{2.14}\\
    f_{\rm{disk}} & = 0.05 \pm^{0.12}_{0.03}
    \end{split}
\end{equation*}
for the fiducial 65~pc distance cut.

\begin{figure}[t!]
    \includegraphics[width=8.5cm]{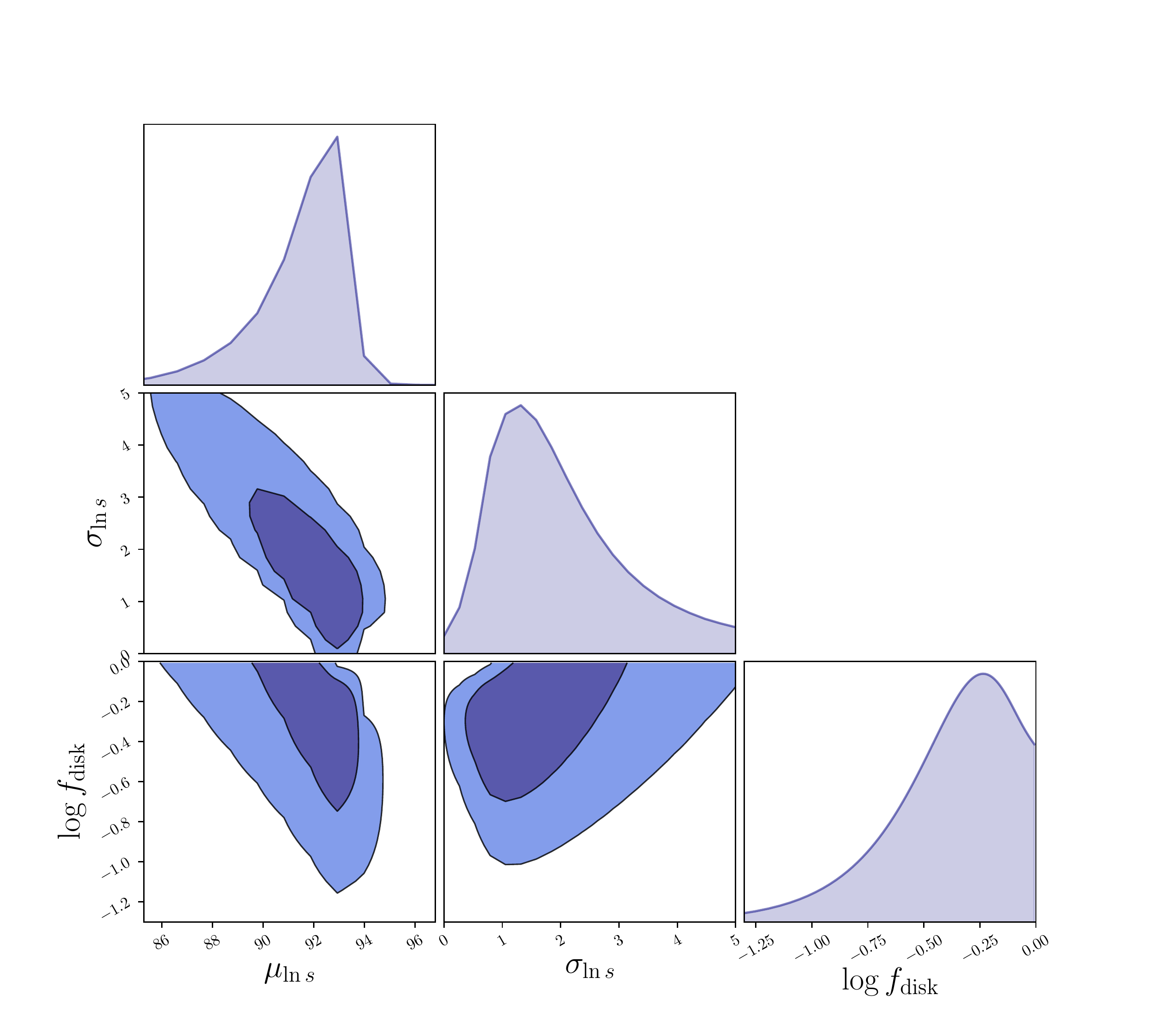}
    \centering
     \caption{Parameter constraints for the log-normal model when the analysis is applied to 75  known debris disks. Our constraint on $f_{\rm disk} = 1$ for this subset of stars is consistent with 1.0, as expected.}\label{fig:75 Known Disks}
\end{figure}

\begin{figure}[h]
    \includegraphics[width=8.5cm]{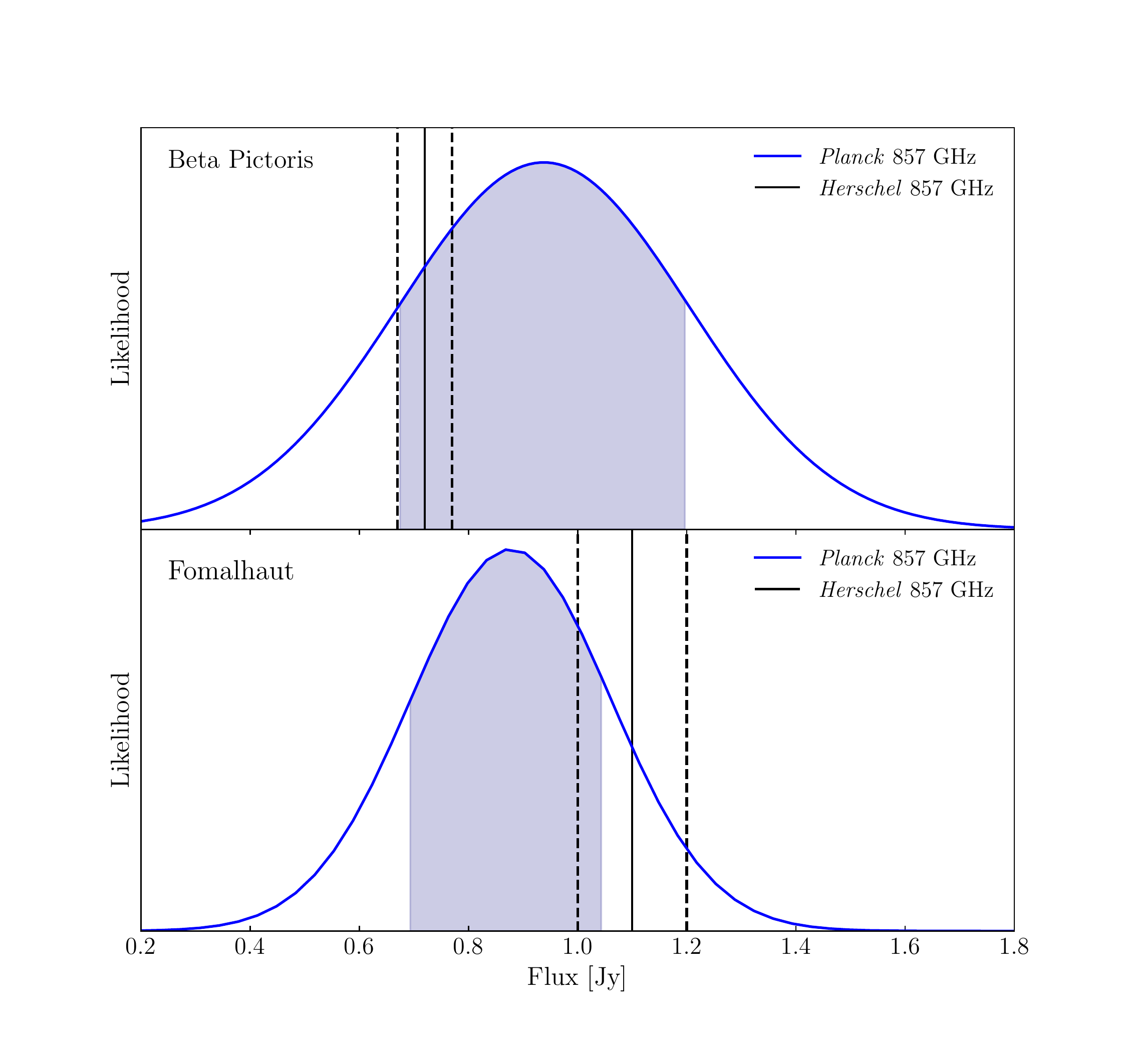}
    \centering
    \caption{Flux measurements of debris disks around Beta Pictoris and Fomalhaut from the {\it Planck} 857~GHz channel with the constant luminosity model. The shaded region represents the 68\% confidence interval, and the solid black lines correspond to measurements of the same stars from {\it Herschel} at 857~GHz. The dashed vertical lines represent the uncertainties on the {\it Herschel} measurements.  }\label{fig:Beta_Pictoris_Fomalhaut}
\end{figure}

\subsection{Log-Normal Luminosity Model Applied to Known Debris Disks}
In this section we center the log-normal model analysis on a catalogue of 75 resolved debris disks from \url{https://www.circumstellardisks.org}. We expect to find measurements consistent with $f_{\rm{disk}} = 1.0$, since each pixel in this sample of stars is known to host a disk.

In Fig.~\ref{fig:75 Known Disks}, we present results for this subset of stars. Indeed, we find constraints consistent with $f_{\rm{disk}}$ = 1.0, $\mu_{\ln{s}} = 92.9 \pm^{2.11}_{5.26}$, and $\sigma_{\ln{s}} = 1.32 \pm^{2.89}_{1.05}$ at 95\% confidence.

\subsection{Measurements around Individual Stars}\label{app: individual stars}
In this section we present the results of our analysis applied to two well studied debris disks systems surrounding Beta Pictoris (19.4~pc) and Fomalhaut (7.7~pc) These disks have extensive observations at 857~GHz with {\it Herschel}, conveniently the same frequency channel used in this analysis. 

We measure the flux from each disk using the constant luminosity model, and compare our measurements to those derived from observation with {\it Herschel} 857~GHz by \citet{Herschel_Fomalhaut} and \citet{2010A&A...518L.133V} for Fomalhaut and Beta Pictoris respectively. Results for this test are shown in Fig.~\ref{fig:Beta_Pictoris_Fomalhaut}. We note that for both stars, the best fit $f_{\rm{disk}}$ is 1.0. Marginalizing over $f_{\rm{disk}}$, we find that our flux measurements are in agreement with {\it Herschel} observations of both disks to within the uncertainties.

\end{document}